# Development of the modified quasichemical model in the distinguishable-pair approximation for multiple compositions of short-range ordering

Kun Wang

*State Key Laboratory of Advanced Special Steel & Shanghai Key Laboratory of Advanced Ferrometallurgy & School of Materials Science and Engineering, Shanghai University, Shanghai 200072, China*

Abstract: A binary solution with short-range ordering (SRO) exhibits characteristic solution thermodynamics. The Modified Quasichemical Model in the Pair Approximation (MQMPA) can effectively capture the thermodynamic features of a binary solution with a onefold SRO. However, once the SRO occurs at multiple compositions in a binary solution, the MQMPA is in principle inconvenient to treat the solution thermodynamics. It usually requires a large number of model parameters to fit the thermodynamic data of such a solution. The present work proposes the Modified Quasichemical Model in the Distinguishable-Pair Approximation (MQMDPA), which is a further improvement of the MQMPA. The MQMDPA can more realistically describe the solution thermodynamics with manifold SROs using fewer model parameters. It benefits from grouping the ordered pairs, which were assumed to be indistinguishable within the MQMPA, into several distinguishable types. Each kind of ordered pair has unique coordination numbers and pair energy, which is responsible for describing one of the SROs at the desired composition and strength. Interestingly, the MQMDPA can be completely transformed into the MQMPA when all kinds of ordered pairs are assigned the same pair energy and coordination numbers. The distinguishable pairs have thus become indistinguishable. Several real liquids with at least two observed SROs were successfully treated by the MQMDPA to demonstrate its effectiveness and reliability.

## 1. Introduction

Accurate thermodynamic predictions of phase diagrams require reliable thermodynamic models for solutions. The Modified Quasichemical Model (MQM) proposed by Pelton et al. [1-4] can be viewed as one of the most trustworthy solution models. The MQM was developed by

Corresponding author.

E-mail address: wangkun0808@shu.edu.cn

making several valuable modifications to the classical quasichemical model of Fowler and Guggenheim [5]. These modifications include (1) permitting the composition of maximum SRO to be freely chosen in a binary system, (2) allowing the coordination numbers to vary with composition, (3) expressing the formation energy of structural entities (pairs or quadruplets) as a polynomial in the entity fractions, (4) extending the model to multicomponent systems, and (5) expanding the model to a two-sublattice framework for reciprocal melts. Most recently, Wang and Chartrand [6] also implemented a modification to the MQM so that the default Gibbs energy of a reciprocal quadruplet can be better defined no matter where the reciprocal SRO occurs.

MQM can circumvent weaknesses that intrinsically exist in the Bragg-Williams Model (BWM) [7], the Associate Solution Model (ASM) [8], and the Ionic Liquid Model (ILM) [9]. The weakness of the BWM is its failure to reproduce the enthalpy and entropy of mixing data in a binary solution with strong SRO. The common weakness of the ASM and ILM is their inability to reduce to the BWM. In addition, both the ternary ASM and ILM are weak with respect to predicting the observed positive deviation from ideality in a ternary solution where strong SROs occur in some of the sub-binary solutions [10]. The MQM is thus very flexible and credible in treating solutions with various configurations.

However, when a binary solution is characteristic of multiple compositions of SRO, the MQMPA is inconvenient and most likely requires a large number of model parameters to treat the solution thermodynamics. This is because the quasichemical model and the modified version [1] can inherently only provide unique ordered pairs to characterize onefold SRO in a binary solution. Realistically, manifold SROs should be spontaneously characterized by multiple types of ordered pairs. To solve this paradox, the MQMPA and the MSM [11] can be coupled to spawn different types of ordered pairs and treat manifold SROs. Wang et al. [12] accordingly proposed the combinatorial model ($K^+$, $Be^{2+}$, $Be_2^{4+}$//$F^-$) to treat the KF-$BeF_2$ liquid wherein the twofold SROs (with the approximate stoichiometry of $K_2BeF_4$ and $K_3BeF_7$) were captured by the K-F-Be and K-F-$Be_2$ pairs (second nearest neighbor) with the respective monomer $Be^{2+}$ and dimer $Be_2^{4+}$ states. Such a treatment actually generated a mixture of $BeF_2$ and $Be_2F_4$ for the pure $BeF_2$ liquid, which resulted in the formation of the pseudoternary solution model (KF-$BeF_2$-$Be_2F_4$) for the binary liquid. This treatment may bring about many uncertain issues, as will be discussed in Section 4.

The present paper makes a further and valuable modification to the MQMPA. The ordered pairs were assumed to be identical in the MQMPA but are distinguished into different types. Each type of ordered pair is indistinguishable and possesses a unique bond energy and coordination numbers. Based on this concept, the MQMDPA is developed to realistically and directly treat a binary solution containing multiple compositions of SRO. Since the MQMDPA can directly provide multiple types of ordered pairs to characterize multiple SROs in a binary solution, it thus avoids those uncertain problems originating from the combinatorial model. Meanwhile, the MQMDPA can reach several limits, which is of great importance to the development of a theoretical model that has not been pursued by the combinatorial model. One of the limiting cases is the complete transformation of the MQMDPA into the MQMPA. This transformation can be triggered if all types of ordered pairs are assigned the same bond energy and coordination numbers. All the model details regarding the MQMDPA (formalisms, limiting cases and internal equilibria) will be displayed in Section 2. Section 3 will select three binary liquid alloys with at least two observed SROs to examine the effectiveness and reliability of the MQMDPA. Section 4 will provide detailed comparisons between the MQMDPA and other solution models, which help reveal uncertain problems arising from the latter in the description of a binary solution containing manifold SROs. Concluding remarks of this paper will be given in Section 5.

## 2. The model

There are two parts in this section. The first part presents the MQMDPA formalism, where all of the external variables (temperature and composition) and internal variables (number of moles of pairs), as well as their interrelations, have been defined in detail. The way to solve the internal variables is provided given the fixed external variables, which leads to the back-calculation of the Gibbs energy. The second part engages in discussions of some important limits that should be reached by a theoretically sound model and can be achieved by the MQMDPA.

**2.1 The formalism**

The Gibbs energy of the A-B solution with $Q$ compositions of maximum short-range ordering can be formulated as,

$$G = (n_A g_A^o + n_B g_B^o) - T\Delta S^{config} + \sum_{k=1}^{Q} n_{AB}^{\langle k \rangle} \left( \frac{\Delta g_{AB}^{\langle k \rangle}}{2} \right) \qquad (1)$$

where $g_A^o$ and $n_A$ are the respective molar Gibbs energy and number of moles of the pure component A ($g_B^o$ and $n_B$ for B), $\Delta S^{config}$ represents the configurational entropy of mixing, $T$ refers to the temperature in Kelvin, and $\Delta g_{AB}^{\langle k \rangle}$ and $n_{AB}^{\langle k \rangle}$ stand for the nonconfigurational Gibbs energy change for the formation of two moles of and the number of moles of the A-B pair in the $k$ type. The first term in Equation (1) is the linear summation of the Gibbs energies from the pure components.

The second term in Equation (1) stands for the Gibbs energy to be contributed from the configurational entropy of mixing. Theoretically, $\Delta S^{config}$ should be given by randomly distributing the (A-A), (B-B) and $Q$ types of (A-B) pairs on three-dimensional lattice sites. However, the entropy of this distribution is unknown since there is no exact solution to the three-dimensional Ising model. Within the framework of the quasichemical model, the exact expression in one dimension is used to approximate the entropy. The approximated entropy is composed of two parts,

$$\Delta S^{config} = \Delta S^{point} + \Delta S^{pair} \qquad (2)$$

where the first and second terms refer to the entropies from the point and pair approximations, respectively. The point-approximation term is shown with the Bragg-Williams formalism as,

$$\Delta S^{point} = -R(n_A \ln X_A + n_B \ln X_B) \qquad (3)$$

where $X_A$ and $X_B$ are the mole fractions of the pure components, and $R$ is the gas constant. The pair-approximation term is expressed by,

$$\Delta S^{pair} = -R \left[ n_{AA} \ln \frac{X_{AA}}{Y_A^2} + n_{BB} \ln \frac{X_{BB}}{Y_B^2} + \sum_{k=1}^{Q} n_{AB}^{\langle k \rangle} \ln \frac{Q X_{AB}^{\langle k \rangle}}{2 Y_A Y_B} \right] \qquad (4)$$

where $Y_A$ is the "coordination-equivalent" fraction of the pure component A ($Y_B$ for B), $n_{AA}$ and $X_{AA}$ are the number of moles of and the mole fraction of the (A-A) pair ($n_{BB}$ and $X_{BB}$ for the (B-B) pair), respectively, and $X_{AB}^{\langle k \rangle}$ is the mole fraction of the $(A-B)^{\langle k \rangle}$ pair.

The last term in Equation (1) stands for the Gibbs energy of formation of $Q$ types of the $(A-B)^{\langle k \rangle}$ pairs, each type with $n_{AB}^{\langle k \rangle}$ mole. The reaction of forming the $(A-B)^{\langle k \rangle}$ pair can be schematically shown as,

$$(A-A) + (B-B) = 2(A-B)^{\langle k \rangle} \qquad \Delta g_{AB}^{\langle k \rangle} \qquad (5)$$

which is similar to the other types. There are in total $Q$ types of such reactions in the A-B solution. All of these represent the first-nearest-neighbor (FNN) pairs. As stated below Equation (1), $\Delta g_{AB}^{\langle k \rangle}$ is the Gibbs energy change for the formation of two moles of $(A - B)^{\langle k \rangle}$ through reaction (5). Similar to the MQMPA, $\Delta g_{AB}^{\langle k \rangle}$ in the MQMDPA can also be expressed as a polynomial in terms of the pair fractions $X_{AA}$ and $X_{BB}$,

$$\Delta g_{AB}^{\langle k \rangle} = \Delta g_{AB}^{\langle k,o \rangle} + \sum_{i \geq 1} g_{AB}^{\langle k,i0 \rangle} X_{AA}^i + \sum_{j \geq 1} g_{AB}^{\langle k,0j \rangle} X_{BB}^j \qquad (6)$$

where $\Delta g_{AB}^{\langle k,o \rangle}$, $g_{AB}^{\langle k,i0 \rangle}$ and $g_{AB}^{\langle k,0j \rangle}$ are the model parameters that can be functions of temperature. $i$ and $j$ are integers exponentializing the pair fractions to perform the fine reproduction of experimental data.

The substance quantities in Equation (1) have the following interrelational equations. The relation between the mole fractions of and the number of moles of the pure components can be given as

$$X_A = \frac{n_A}{n_A + n_B} = 1 - X_B \qquad (7)$$

for the binary A-B solution. Let $Z_A$ and $Z_B$ be the overall coordination numbers of A and B. Then, the mass-balance relations between the pure components and the related pairs are derived as

$$Z_A n_A = 2n_{AA} + \sum_{k=1}^{Q} n_{AB}^{\langle k \rangle} \qquad (8)$$

$$Z_B n_B = 2n_{BB} + \sum_{k=1}^{Q} n_{AB}^{\langle k \rangle} \qquad (9)$$

From combining equations (8-9), the following equation results:

$$\frac{Z_A n_A + Z_B n_B}{2} = n_{AA} + n_{BB} + \sum_{k=1}^{Q} n_{AB}^{\langle k \rangle} \qquad (10)$$

where the term on the right-hand side represents the total number of all pairs in the A-B solution. The pair fraction of $(A - B)^{\langle k \rangle}$ is defined as,

$$X_{AB}^{\langle k \rangle} = \frac{n_{AB}^{\langle k \rangle}}{n_{AA} + n_{BB} + \sum_{k=1}^{Q} n_{AB}^{\langle k \rangle}} \qquad (11)$$

which is similar to the definitions of the (A-A) and (B-B) pairs. The coordination-equivalent fractions are defined as

$$Y_A = \frac{Z_A n_A}{Z_A n_A + Z_B n_B} = \frac{Z_A X_A}{Z_A X_A + Z_B X_B} = 1 - Y_B \qquad (12)$$

for the binary solution. Combining equations (8-12), the relations between the "coordination-equivalent" fractions and the relative pair fractions are derived as

$$Y_A = X_{AA} + \frac{\sum_{k=1}^{Q} X_{AB}^{\langle k \rangle}}{2} \tag{13}$$

$$Y_B = X_{BB} + \frac{\sum_{k=1}^{Q} X_{AB}^{\langle k \rangle}}{2} \tag{14}$$

For the earlier versions [13-14] of the MQMPA, $Z_A$ and $Z_B$ are constant values and independent of composition. However, immutable coordination numbers bring a number of drawbacks, and they have thus been modified to vary with composition in the current MQMPA [1]. The MQMDPA is developed by also adopting this modification, which generates equations (15-16),

$$\frac{1}{Z_A} = \frac{\frac{2n_{AA}}{Z_{AA}^A} + \sum_{k=1}^{Q} \frac{n_{AB}^{\langle k \rangle}}{Z_{AB}^{A\ \langle k \rangle}}}{2n_{AA} + n_{AB}} \tag{15}$$

$$\frac{1}{Z_B} = \frac{\frac{2n_{BB}}{Z_{BB}^B} + \sum_{k=1}^{Q} \frac{n_{AB}^{\langle k \rangle}}{Z_{AB}^{B\ \langle k \rangle}}}{2n_{BB} + n_{AB}} \tag{16}$$

where $Z_{AA}^A$ and $Z_{AB}^{A\ \langle k \rangle}$ are the values of $Z_A$ when all nearest neighbors of an A are As, and when all nearest neighbors of an A are Bs forming $(A-B)^{\langle k \rangle}$, and where $Z_{BB}^B$ and $Z_{AB}^{B\ \langle k \rangle}$ are defined similarly. The $k$th composition of the maximum SRO is determined by the ratio $(Z_{AB}^{A\ \langle k \rangle}/Z_{AB}^{B\ \langle k \rangle})$. Substituting equations (15-16) into equations (8-9) yields

$$n_A = \frac{2n_{AA}}{Z_{AA}^A} + \sum_{k=1}^{Q} \frac{n_{AB}^{\langle k \rangle}}{Z_{AB}^{A\ \langle k \rangle}} \tag{17}$$

$$n_B = \frac{2n_{BB}}{Z_{BB}^B} + \sum_{k=1}^{Q} \frac{n_{AB}^{\langle k \rangle}}{Z_{AB}^{B\ \langle k \rangle}} \tag{18}$$

Equations (12-14) remain unchanged. As pointed out by Pelton et al. [1], the composition dependence of equations (15-16) is chosen because it results in the simple relationships of equations (17-18). This simplifies subsequent calculations, and allows the direct comparison between the MQMDPA and the ASM by analyzing their expressions, as elaborated in Section 4.3.

To date, all substance quantities, as well as their interrelations, have already been defined. The equilibrium distribution of pairs in the solution can be obtained by minimizing $G$ of Equation (1) as

$$\left(\frac{\partial G}{\partial n_{AB}^{\langle k \rangle}}\right)_{T, n_A, n_B, n_{AB}^{\langle k' \rangle}|_{k \neq k'}} = 0 \tag{19}$$

at fixed $n_A$ and $n_B$, subjected to the constraints of equations (8-9). This gives

$$\left(\frac{X_{AA}}{Y_A^2}\right)^{C_{AA}^{\langle k \rangle}} \left(\frac{X_{BB}}{Y_B^2}\right)^{C_{BB}^{\langle k \rangle}} \left(\frac{Q X_{AB}^{\langle k \rangle}}{2 Y_A Y_B}\right) = \exp\left(\frac{-\Delta g_{AB}^{\langle k \rangle}}{2RT}\right) \tag{20}$$

where

$$C_{AA}^{\langle k \rangle} = \frac{-Z_{AA}^A}{2 Z_{AB}^{A\ \langle k \rangle}} \tag{21}$$

$$C_{BB}^{\langle k \rangle} = \frac{-Z_{BB}^B}{2 Z_{AB}^{B\ \langle k \rangle}} \tag{22}$$

There are $Q$ equations as Equation (19), which also generates $Q$ equations as Equation (20). Equation (20) is the "equilibrium constant" for the quasichemical reaction (5). For given values of $\Delta g_{AB}^{\langle k \rangle}$, the solution of Equation (20) together with equations (11, 17-18) gives $n_{AA}$, $n_{BB}$ and $n_{AB}^{\langle k \rangle}$, which can then be substituted into Equation (1) to obtain the Gibbs energy and its thermodynamic relatives at the defined $n_A$ and $n_B$.

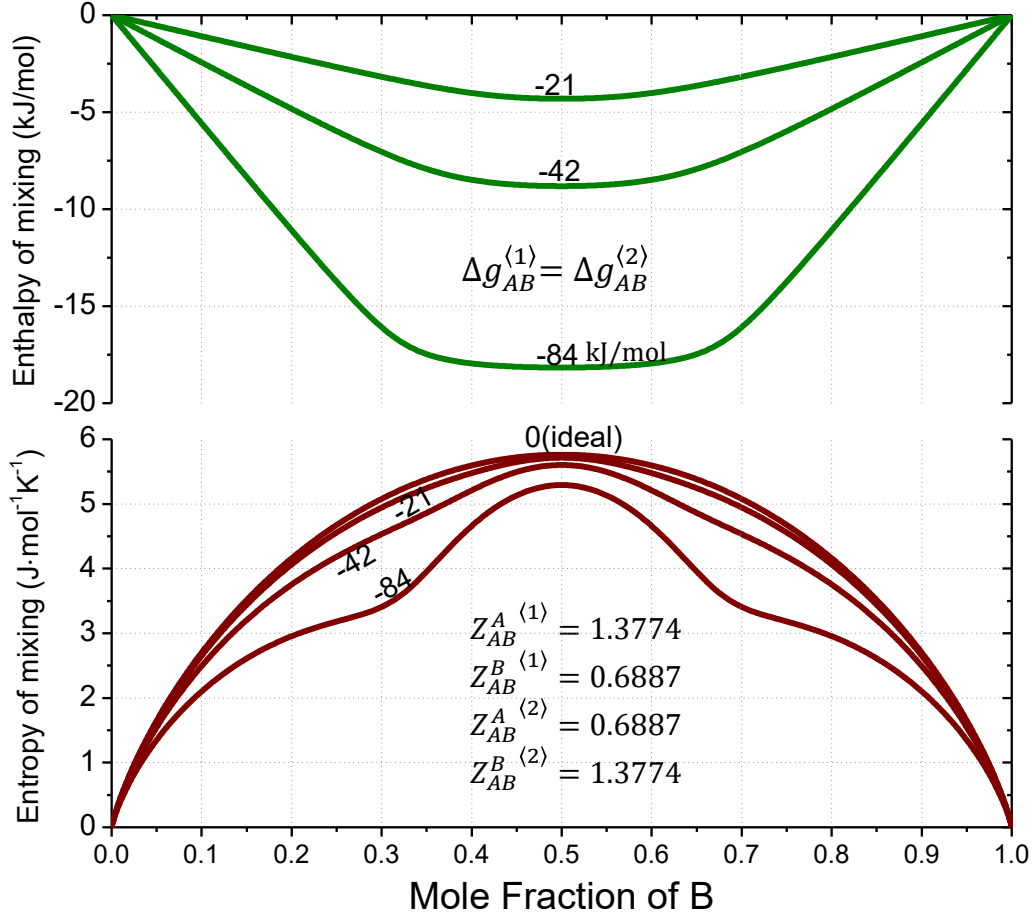

Fig. 1 Molar enthalpy and entropy of mixing for an arbitrary A-B system calculated at 1000 °C from the MQMDPA for twofold SROs with $\Delta g_{AB}^{\langle 1 \rangle} = \Delta g_{AB}^{\langle 2 \rangle} = 0, -21, -42, -84 \ kJ/mol$ and the coordination numbers defined in Equation (24)

To capture two SROs in the A-B solution, the MQMDPA with $Q=2$ is used. According to equations (1-4), the Gibbs energy function then reduces as

$$G = n_A g_A^o + n_B g_B^o + RT \left( n_A \ln X_A + n_B \ln X_B + n_{AA} \ln \frac{X_{AA}}{Y_A^2} + n_{BB} \ln \frac{X_{BB}}{Y_B^2} + n_{AB}^{\langle 1 \rangle} \ln \frac{X_{AB}^{\langle 1 \rangle}}{Y_A Y_B} + n_{AB}^{\langle 2 \rangle} \ln \frac{X_{AB}^{\langle 2 \rangle}}{Y_A Y_B} \right) + n_{AB}^{\langle 1 \rangle} \frac{\Delta g_{AB}^{\langle 1 \rangle}}{2} + n_{AB}^{\langle 2 \rangle} \frac{\Delta g_{AB}^{\langle 2 \rangle}}{2} \quad (23)$$

Based on Equation (23), the enthalpy and entropy of mixing for an arbitrary system were calculated at 1000 °C with various values of $\Delta g_{AB}^{\langle 1 \rangle}$ and $\Delta g_{AB}^{\langle 2 \rangle}$. The calculated results are shown in Fig.1. When both $\Delta g_{AB}^{\langle 1 \rangle}$ and $\Delta g_{AB}^{\langle 2 \rangle}$ become progressively more negative with assigned $Z_{AB}^{A\ \langle 1 \rangle}/Z_{AB}^{B\ \langle 1 \rangle} = 2/3$ and $Z_{AB}^{A\ \langle 2 \rangle}/Z_{AB}^{B\ \langle 2 \rangle} = 1/3$, the formations of the $(A-B)^{\langle 1 \rangle}$ and $(A-B)^{\langle 2 \rangle}$ pairs progressively proceed through reaction (5). The two SROs thus occur at $X_B=1/3$ and $X_B=2/3$. As illustrated in Fig. 1, the calculated enthalpy of mixing versus composition displays a sunken plain within the two SRO compositions, and the calculated configurational entropy also shows two concavities around the two compositions. The solution within the two SRO compositions will merely contain the $(A-B)^{\langle 1 \rangle}$ and $(A-B)^{\langle 2 \rangle}$ pairs if $\Delta g_{AB}^{\langle 1 \rangle}$ and $\Delta g_{AB}^{\langle 2 \rangle}$ approach infinite negativity. Most likely, there exists a miscibility gap within the solution zone.

Since the quasichemical model is built based upon a one-dimensional Ising model, the appearance of negative configurational entropy should always be considered if it is applied to the real solutions in three dimensions with very negative $\Delta g_{AB}$. This is also true for the present MQMDPA. To avoid negative configurational entropy, the coordination number should be defined as

$$z_B = \left[\ln\left(1 + \frac{X_A}{X_B}\right) + \frac{X_A}{X_B}\ln\left(1 + \frac{X_B}{X_A}\right)\right]/\ln(2Q) \qquad (24)$$

where the mole fraction ($X_A$ or $X_B$) is the composition where the maximum SRO occurs. The configurational entropy can become zero using very negative $\Delta g_{AB}^{\langle k \rangle}$ and the coordination numbers defined in Equation (24). In this case, there are only $(A-B)^{\langle k \rangle}$ pairs, and the solution exhibits virtually complete SRO at the composition. According to Equation (24), the coordination numbers displayed in Fig. 1 are used for all the thermodynamic calculations.

## 2.2 Limiting cases

The Bragg-Williams approximation is the first limit that the MQMDPA can reach given that all the $\Delta g_{AB}^{\langle k \rangle}$ values approach zero regardless of how all the coordination numbers are defined. If all the $\Delta g_{AB}^{\langle k \rangle}$ values are small enough, random distributions of A and B can become true over the quasi-lattice sites. This gives $X_{AA} = Y_A^2$ and $X_{BB} = Y_B^2$, which are then substituted into Equation (20) to yield $X_{AB}^{\langle k \rangle} = 2Y_A Y_B/Q$. Substituting these into Equation (4) causes $\Delta S^{pair}$ to be nil, and $\Delta S^{config}$ is thus equivalent to the ideal entropy of mixing. In this case, Equation (1) reduces to the expression of the BWM,

$$G = (n_A g_A^o + n_B g_B^o) + RT(n_A \ln X_A + n_B \ln X_B) + \sum_{k=1}^{Q} n_{AB}^{\langle k \rangle}\left(\frac{\Delta g_{AB}^{\langle k \rangle}}{2}\right) \qquad (24)$$

where all the $n_{AB}^{\langle k \rangle}$ values are closely equivalent. According to Equation (10), $n_{AB}^{\langle k \rangle}$ can be expressed as

$$n_{AB}^{\langle k \rangle} = X_{AB}^{\langle k \rangle}\left(n_{AA} + n_{BB} + \sum_{k=1}^{Q} n_{AB}^{\langle k \rangle}\right) = \frac{2Y_A Y_B}{Q}\frac{Z_A n_A + Z_B n_B}{2} \qquad (25)$$

which is substituted into Equation (24) to form

$$G = (n_A g_A^o + n_B g_B^o) + RT(n_A \ln X_A + n_B \ln X_B) + (Z_A n_A + Z_B n_B)Y_A Y_B \sum_{k=1}^{Q} \frac{\Delta g_{AB}^{\langle k \rangle}}{2Q} \quad (26)$$

Per mole of solution ($n_A + n_B = 1$), Equation (26) can be changed as

$$g = (X_A g_A^o + X_B g_B^o) + RT(X_A \ln X_A + X_B \ln X_B) + (Z_A X_A + Z_B X_B)Y_A Y_B \sum_{k=1}^{Q} \frac{\Delta g_{AB}^{\langle k \rangle}}{2Q} \quad (27)$$

If all the $\Delta g_{AB}^{\langle k \rangle}$ values are equal to zero, Equation (27) becomes the ideal solution expression. When $Z_A = Z_B$, then $Y_A = X_A$ and $Y_B = X_B$. If all the $\Delta g_{AB}^{\langle k \rangle}$ expressions are independent of composition, Equation (27) reduces to the well-known regular solution expression. If any $\Delta g_{AB}^{\langle k \rangle}$ is expanded as a polynomial as in Equation (6), Equation (27) can transform into the substitutional solution model widely used for binary solutions with no strong SRO. The last term in Equation (27) is then called the excess molar Gibbs energy and is commonly expressed as a polynomial in the mole fractions. The first limiting case can be clearly illustrated in Fig. 1.

The MQMDPA can approach the second limit in the following scenarios. Suppose two SROs occur in the A-B solution. According to Equation (20), the equilibrium constants for the two corresponding quasichemical reactions are given,

$$\left(\frac{X_{AA}}{Y_A^2}\right)^{C_{AA}^{\langle 1 \rangle}} \left(\frac{X_{BB}}{Y_B^2}\right)^{C_{BB}^{\langle 1 \rangle}} \left(\frac{X_{AB}^{\langle 1 \rangle}}{Y_A Y_B}\right) = \exp\left(\frac{-\Delta g_{AB}^{\langle 1 \rangle}}{2RT}\right) \quad (28)$$

$$\left(\frac{X_{AA}}{Y_A^2}\right)^{C_{AA}^{\langle 2 \rangle}} \left(\frac{X_{BB}}{Y_B^2}\right)^{C_{BB}^{\langle 2 \rangle}} \left(\frac{X_{AB}^{\langle 2 \rangle}}{Y_A Y_B}\right) = \exp\left(\frac{-\Delta g_{AB}^{\langle 2 \rangle}}{2RT}\right) \quad (29)$$

where $C_{AA}^{\langle k \rangle}$ and $C_{BB}^{\langle k \rangle}$ ($k$=1, 2) are calculated through equations (21-22) using the coordination numbers defined above. The above definition gives the following relation:

$$C_{AA}^{\langle 2 \rangle} - C_{AA}^{\langle 1 \rangle} + C_{BB}^{\langle 2 \rangle} - C_{BB}^{\langle 1 \rangle} = 0 \quad (30)$$

Equation (28) divided by Equation (29) yields

$$X_{AA}^{(C_{AA}^{\langle 1 \rangle} - C_{AA}^{\langle 2 \rangle})} X_{BB}^{(C_{BB}^{\langle 1 \rangle} - C_{BB}^{\langle 2 \rangle})} Y_A^{2(C_{AA}^{\langle 2 \rangle} - C_{AA}^{\langle 1 \rangle})} Y_B^{2(C_{BB}^{\langle 2 \rangle} - C_{BB}^{\langle 1 \rangle})} \left(\frac{X_{AB}^{\langle 1 \rangle}}{X_{AB}^{\langle 2 \rangle}}\right) = \exp\left(\frac{\Delta g_{AB}^{\langle 2 \rangle} - \Delta g_{AB}^{\langle 1 \rangle}}{2RT}\right) \quad (31)$$

This can be simplified as

$$\frac{X_{AB}^{\langle 1 \rangle}}{X_{AB}^{\langle 2 \rangle}} = \exp\left(\frac{-\Delta g_{AB}^{\langle 1 \rangle}}{2RT}\right) \quad (32)$$

with Equation (30) and $\Delta g_{AB}^{\langle 2 \rangle} = 0$. When $\Delta g_{AB}^{\langle 1 \rangle} = -\infty$, Equation (32) gives $X_{AB}^{\langle 2 \rangle} = 0$. In this case, the second SRO is extinct, and the solution merely has a onefold SRO. This limiting case can be clearly illustrated in Fig. 2 by the entropy of mixing and the equilibrium distribution of pairs in the A-B solution. In fact, the $(A-B)^{\langle 1 \rangle}$ pairs dominate at the expense of the $(A-B)^{\langle 2 \rangle}$ pairs in the solution. This is because when $\Delta g_{AB}^{\langle 1 \rangle}$ becomes progressively more negative and $\Delta g_{AB}^{\langle 2 \rangle}$ is set to be nil, the quasichemical reaction (5) for $k=1$ will be remarkably shifted to the right, while driving the reaction for $k=2$ to be correspondingly shifted to the left.

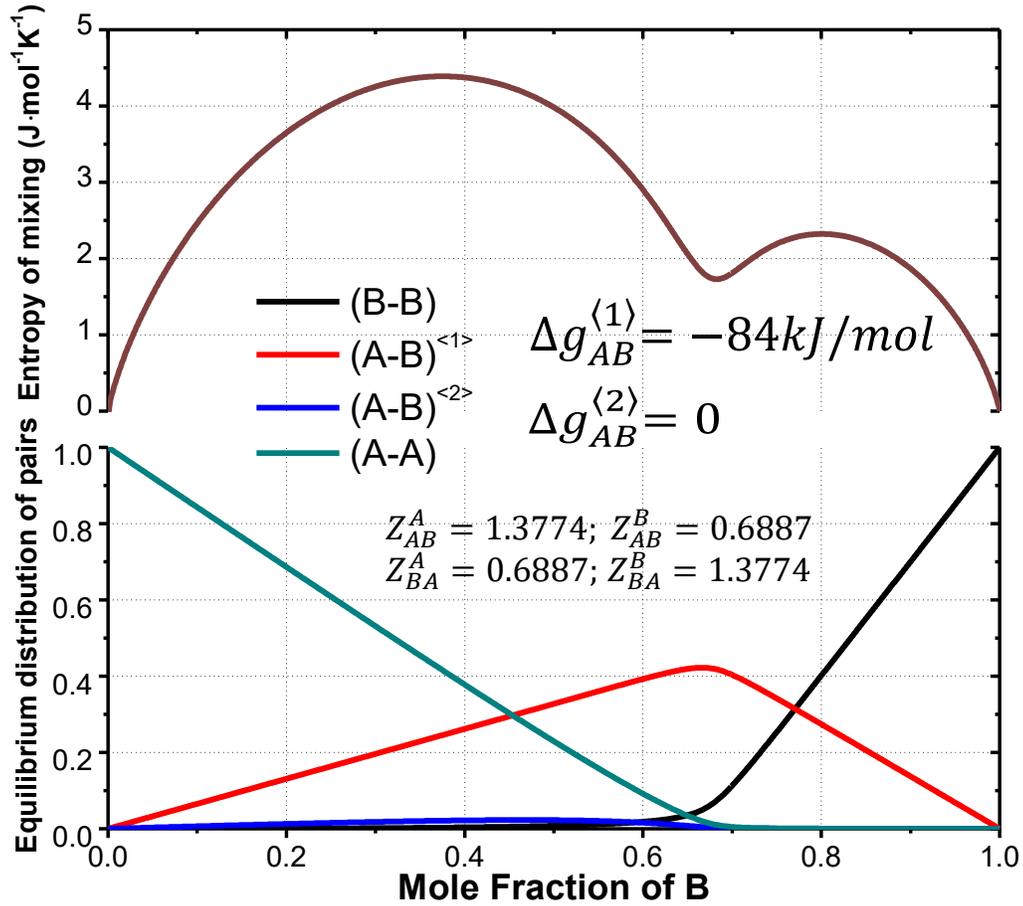

Fig. 2 Molar entropy of mixing and equilibrium distribution of pairs for an arbitrary A-B system calculated at 1000 °C from the MQMDPA for twofold SROs

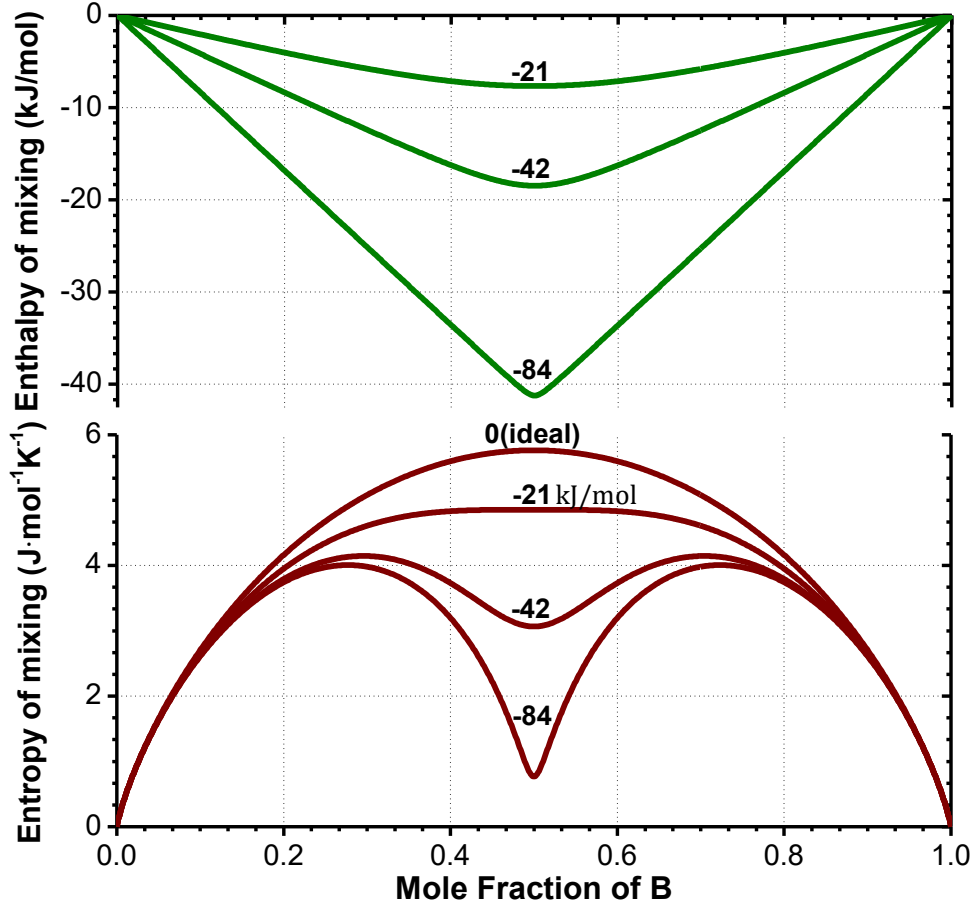

Fig. 3 Molar enthalpy and entropy of mixing for an arbitrary A-B system calculated at 1000 °C from the MQMDPA for twofold SROs with Z=2 and $\Delta g_{AB}^{\langle 1 \rangle} = \Delta g_{AB}^{\langle 2 \rangle} = 0, -21, -42, -84$ kJ/mol

The last and most important limit that the MQMDPA can reach is regarding the complete transformation from the MQMDPA into the MQMPA. The transformation takes place when all types of $(A - B)^{\langle k \rangle}$ have unique bond energy and coordination numbers. This leads to the same content for each type of $(A - B)^{\langle k \rangle}$ in the solution. The following two relations are thus built:

$$n_{AB} = Q n_{AB}^{\langle k \rangle} \tag{33}$$

$$X_{AB} = Q X_{AB}^{\langle k \rangle} \tag{34}$$

where $n_{AB}$ and $X_{AB}$ are the total number of moles of and the mole fraction of all the (A-B) pairs, respectively. Substituting Equations (33-34) into Equation (1) gives

$$G = n_A g_A^o + n_B g_B^o + RT(n_A \ln X_A + n_B \ln X_B) + RT(n_{AA} \ln \frac{X_{AA}}{Y_A^2} + n_{BB} \ln \frac{X_{BB}}{Y_B^2}$$

$$+ n_{AB} \ln \frac{X_{AB}}{2 Y_A Y_B}) + n_{AB} \left(\frac{\Delta g_{AB}}{2}\right) \tag{35}$$

where $\Delta g_{AB}$ has replaced $\Delta g_{AB}^{\langle k \rangle}$ since all types of $(A-B)^{\langle k \rangle}$ have the same bond energy. The distinguishable $(A-B)^{\langle k \rangle}$ pairs have become indistinguishable. Fig.3 shows the enthalpy and entropy of mixing for an arbitrary A-B system calculated by MQMDPA ($Q$=2) with the same bond energy and coordination numbers. It is obvious that the calculated results are identical to those from the MQMPA under the same calculation conditions. In Section 3, three binary liquid alloys with at least two observed SROs are selected to examine the effectiveness and reliability of the MQMDPA.

## 3. Case studies

Short-range ordering widely exists in liquids wherein the alloying atoms have large differences in electronegativity, such as alloys of alkali metals with group III, IV and V metals. Electrons from the former metals can be transferred to the latter ones to form nonmetallic ionic species, and SROs thus occur. The study of statistical mechanics has shown that the structural information regarding molten alloys can be extracted from their thermodynamic properties [15-16]. A good indication of ordering in a solution can be given by the Excess Stability Function (ESF) defined by Darken [17] as,

$$\text{ES} = \left(\frac{\partial^2 G^E}{\partial X_i^2}\right)_T = \frac{RT}{1-X_i}\left(\frac{\partial \ln \gamma_i}{X_i}\right)_T \tag{36}$$

where $G^E$ is the excess Gibbs energy of the solution, and $\gamma_i$ is the activity coefficient of component $i$. Analyses of the measured thermodynamic data using the Darken's excess stability clearly show that there are at least twofold SROs in the Bi-K, Bi-Rb and Na-Pb liquids. For simplicity, the MQMDPA with $Q$=2 is employed to treat these liquids, and their data are correctly adopted to examine the effectiveness and reliability of the model.

Table 1 The coordination numbers and model parameters for the selected liquids ($\Delta g$ in kJ/mol and T in Kelvin)

| Parameters \ Systems | Bi-K | Bi-Rb | Na-Pb |
|---|---|---|---|
| Coordination numbers | $Z_{BiK}^{Bi\ (1)} = 1.5$  $Z_{BiK}^{K\ (1)} = 0.5$ | $Z_{BiRb}^{Bi\ (1)} = 1.5$  $Z_{BiRb}^{Rb\ (1)} = 0.5$ | $Z_{PbNa}^{Pb\ (1)} = 1.0$  $Z_{PbNa}^{Na\ (1)} = 0.25$ |
| | $Z_{BiK}^{Bi\ (2)} = 4$  $Z_{BiK}^{K\ (2)} = 4$ | $Z_{BiRb}^{Bi\ (2)} = 4$  $Z_{BiRb}^{Rb\ (2)} = 4$ | $Z_{PbNa}^{Pb\ (2)} = 5.2$  $Z_{PbNa}^{Na\ (2)} = 4.7$ |
| Model parameters | $\Delta g_{BiK}^{(1)} = -234000 + 105.4T$ | $\Delta g_{BiRb}^{(1)} = -219000 + 100.4T$ | $\Delta g_{NaPb}^{(1)} = -150000 + 105T$ |
| | $\Delta g_{BiK}^{(2)} = -43000 - 3.0T$ | $\Delta g_{BiRb}^{(2)} = -42000 - 4.2T$ | $\Delta g_{NaPb}^{(2)} = -42000 - 4.2T$ |

Table 1 lists all the coordination numbers and model parameters required to describe the Bi-K, Bi-Rb and Na-Pb liquids. It is presented that all the liquids can be well described by the present MQMDPA with much fewer model parameters independent of composition. The MQMPA can employ a large number of model parameters to fit the Gibbs energies and their first derivatives versus composition; it is hardly able to capture the feature of each two-peak ESF curve. The ASM, ILM and combinatorial model can use many model parameters to reproduce the two peaks on the ESF curve but pose other uncertain problems. These uncertain issues will be discussed in Section 4.

**3.1 The Bi-K system**

The thermodynamic properties of the Bi-K liquid were determined over the temperature range of 703 to 933 K from 0 to 80 at.% K by EMF measurements [18]. The obtained integral Gibbs energies, enthalpies, and entropies are very negative. The ESF directly calculated from the scatter data shows two peaks near 50 and 75 at.% K on the curve, which can be interpreted as an indication of maximum structural ordering at the two compositions. According to the ESF, Niu et al. [19] used the ASM with two associates of BiK and $BiK_3$ to describe the Bi-K liquid, and twenty-two model parameters were needed. However, the large number of model parameters used to describe the liquid phase may cause unreasonable extrapolation into higher-order systems. Later, Cao et al. [20] employed the MQMPA to describe the liquid phase with the assumption of a onefold SRO near $BiK_3$ and twelve model parameters in addition to the coordination numbers. The present study employs MQMDPA to describe the Bi-K liquid assuming that the two compositions of maximum SRO possess the stoichiometry of BiK and $BiK_3$. As shown in Table 1, apart from the coordination numbers, only four model parameters independent of composition are required to be able to describe all the experimental data.

Figs. 4-6 present the calculated Gibbs energy of mixing, enthalpy of mixing, and entropy of mixing for the liquid phase at 873 K compared with the experimental data [18]. The calculated results using the MQMDPA are in quite good agreement with the experimental data. The calculated activity coefficient of K versus $X_K$ at 873 K, as shown in Fig. 7, can also effectively reproduce the experimental data. It is obvious that the curve for the activity coefficient versus composition exhibits striking and weak steepness at approximately 75 at.% and 50 at.% K, indicating the formation of strong and weak SROs for the melt configuration, respectively. The two SROs can be more clearly observed from the calculated ESF curve, as presented in Fig. 8. Fine tuning of the model parameters and coordination numbers could be conducted to force the calculated peaks to be excellently consistent with the data points; however, it is meaningless to conduct this adjustment since the ESF is the second derivative of excess Gibbs energy and is very sensitive to its accuracy.

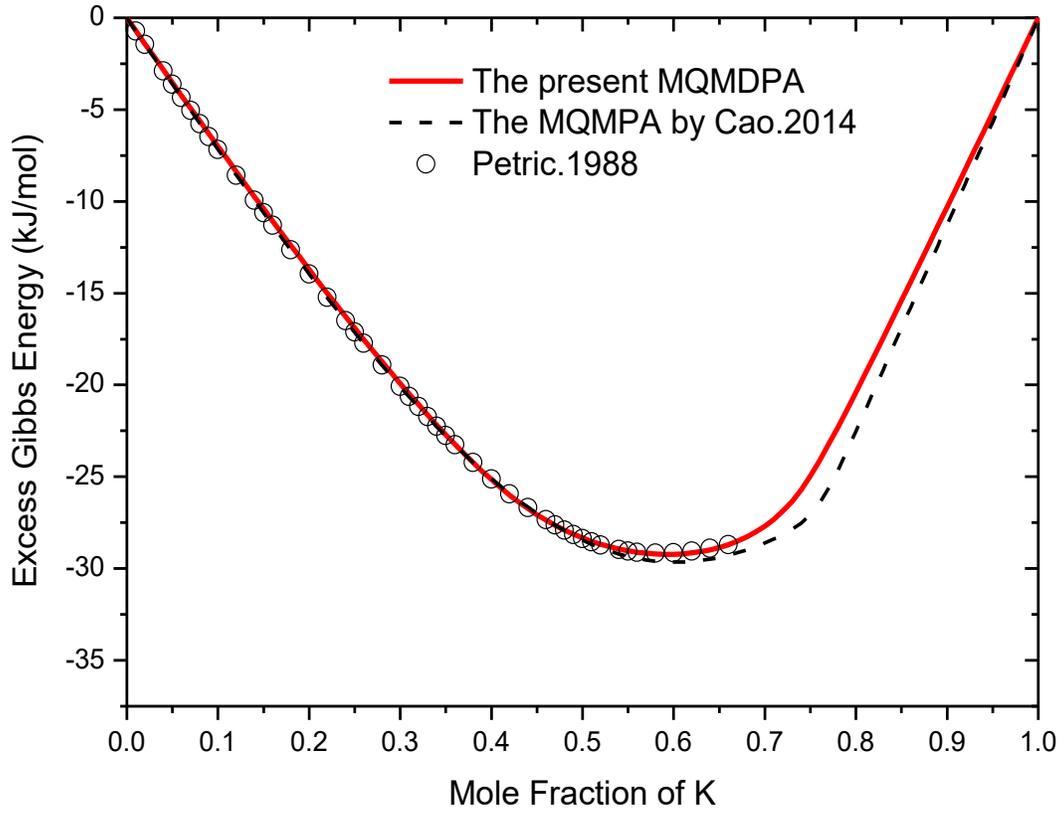

Fig. 4. Calculated Gibbs energy of mixing for the Bi-K liquid at 873 K compared with the experimental data [18]

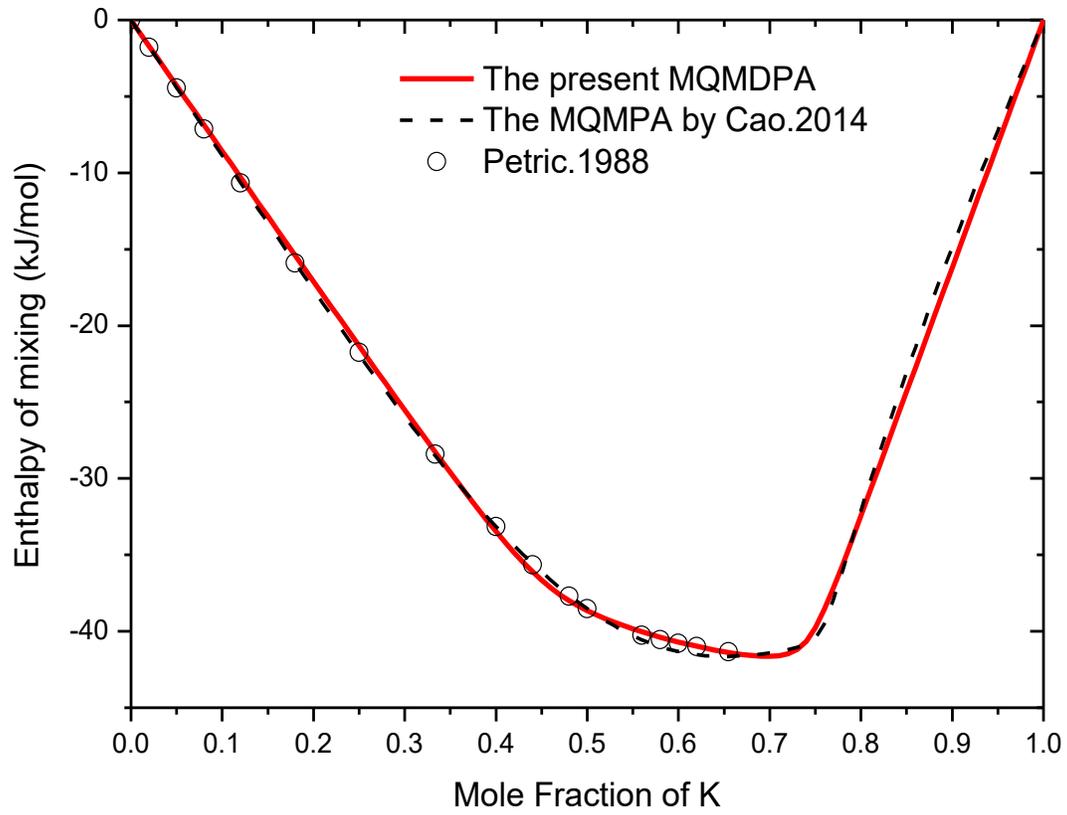

Fig. 5. Calculated enthalpy of mixing for the Bi-K liquid at 873 K compared with the experimental data [18]

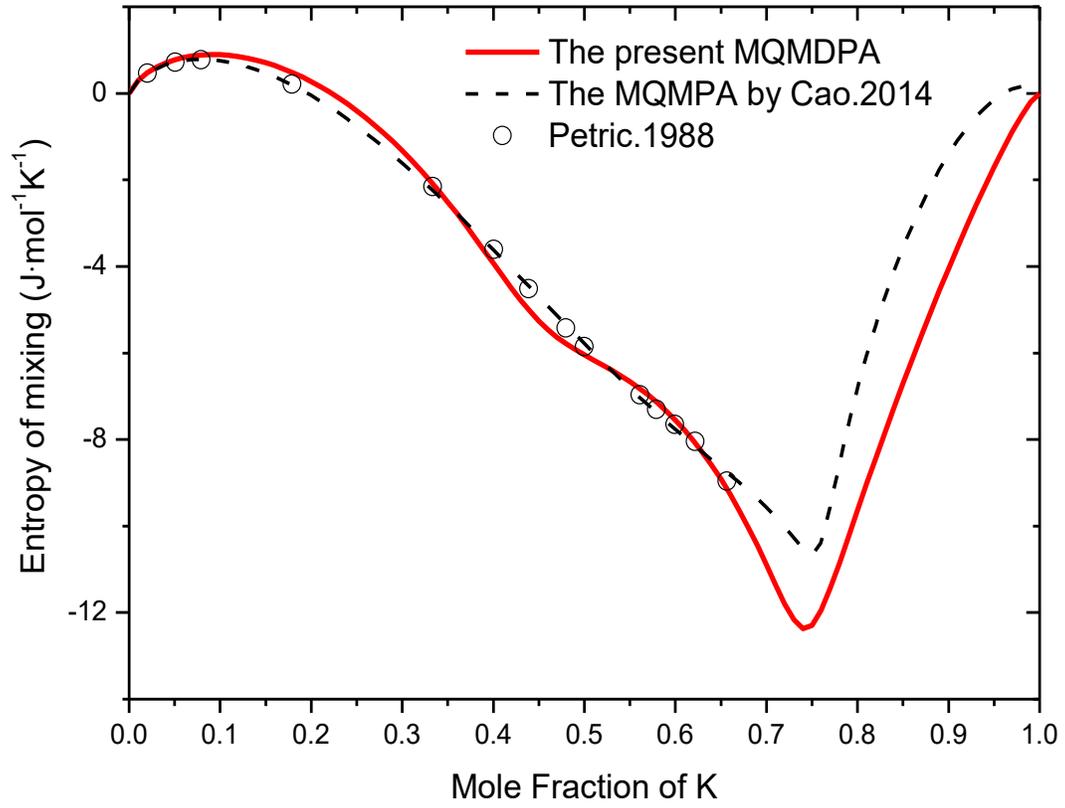

Fig. 6. Calculated entropy of mixing for the Bi-K liquid at 873 K compared with the experimental data [18]

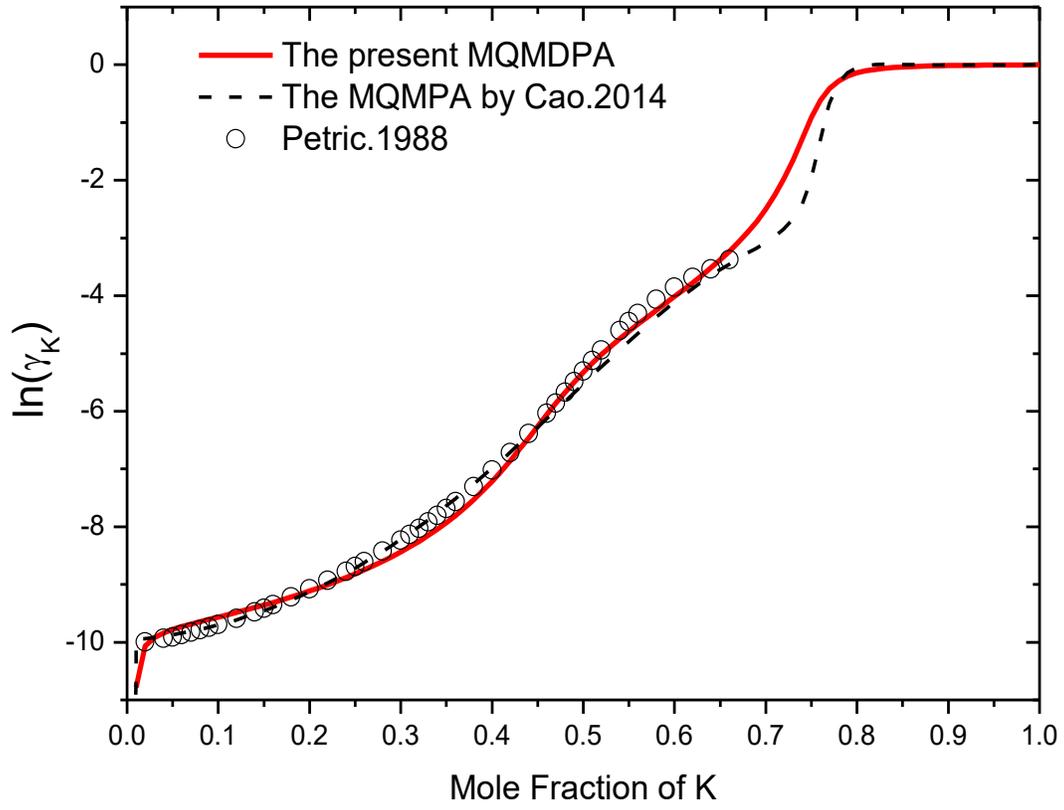

Fig. 7. Calculated activity coefficient of K in the Bi-K liquid at 873 K compared with the experimental data [18]

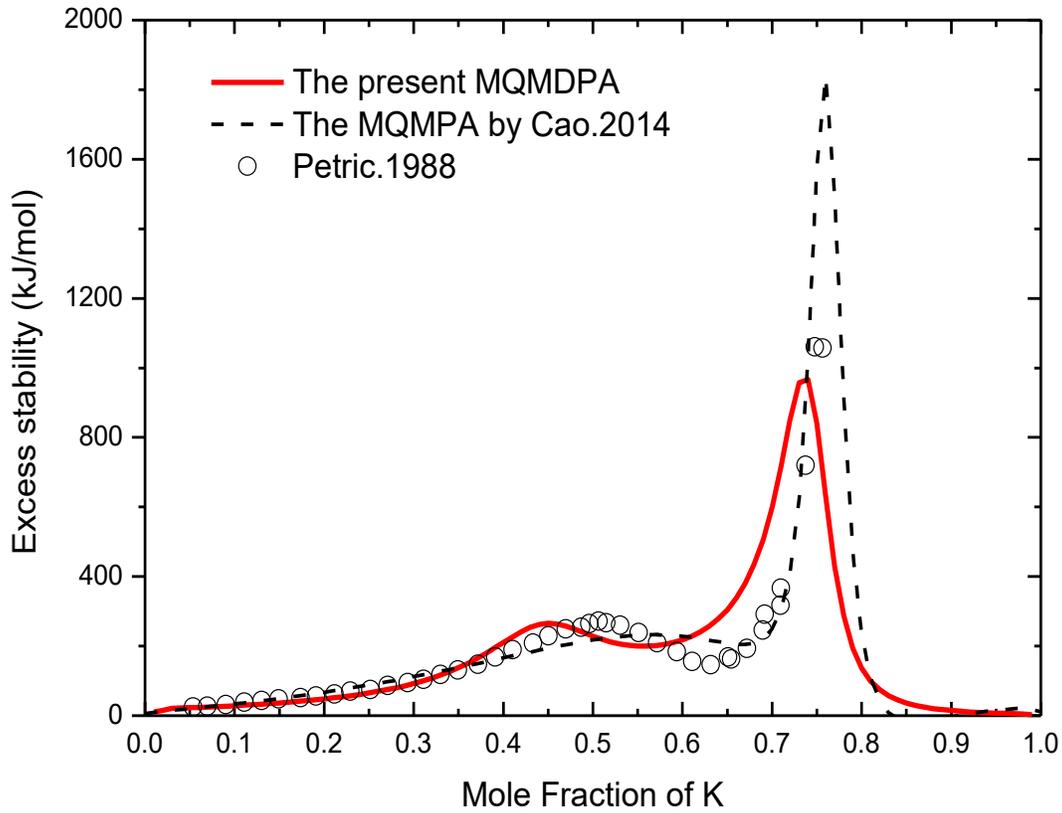

Fig. 8. Calculated excess stability function in the Bi-K liquid at 873 K compared with the experimental data [18]

## 3.2 The Bi-Rb system

The thermodynamic properties of the Bi-Rb liquid were determined over the temperature range from 703 to 933 K and the composition range from 0 to 80 at.% Rb by EMF measurements [21]. The activities of elements, the integral Gibbs energies, enthalpies and entropies of the liquid at 873 K were provided. The calculated ESF curve from the measured scatter data also shows two peaks near 50 and 75 at.% Rb, indicating that two SROs exist. According to the ESF, Liu et al. [22] employed the ASM with two associates of BiRb and $BiRb_3$ to describe the Bi-Rb liquid, and eighteen model parameters were needed. Many model parameters usually cause uncertainties when they are extended to multicomponent solutions. Cao et al. [20] also modeled the Bi-Rb liquid by using the MQMPA with the assumption of onefold SRO near $BiRb_3$ and twelve model parameters in addition to the coordination numbers. This work employs MQMDPA to treat Bi-Rb liquid, assuming that the two compositions of maximum SRO are located at the stoichiometry of BiRb and $BiRb_3$. Apart from the coordination numbers, four model parameters independent of composition as listed in Table 1, are sufficient to fit all the experimental data.

Figs. 9-11 display the calculated Gibbs energy of mixing, enthalpy of mixing, and entropy of mixing for the Bi-Rb liquid at 873 K compared with the experimental data [21]. The images show that the calculated results using the MQMDPA agree well with the experimental data. The calculated activity coefficient of Rb versus $X_{Rb}$ at 873K, as shown in Fig.12, exhibits fine consistency with the experimental data. It is evident that the curve for the activity coefficient versus composition displays striking and weak steepness at approximately 75 at.% and 50 at.% Rb, manifesting the formation of strong and weak SROs in the melt, respectively. The two SROs can also be clearly demonstrated from the calculated ESF curve as presented in Fig. 13. Fine tuning of the model parameters and coordination numbers can be performed to impel the calculated peaks to be excellently consistent with the data points. This makes no practical sense due to the experimental accuracy and the sensitivity in obtaining the ESF using the second derivative of excess Gibbs energy.

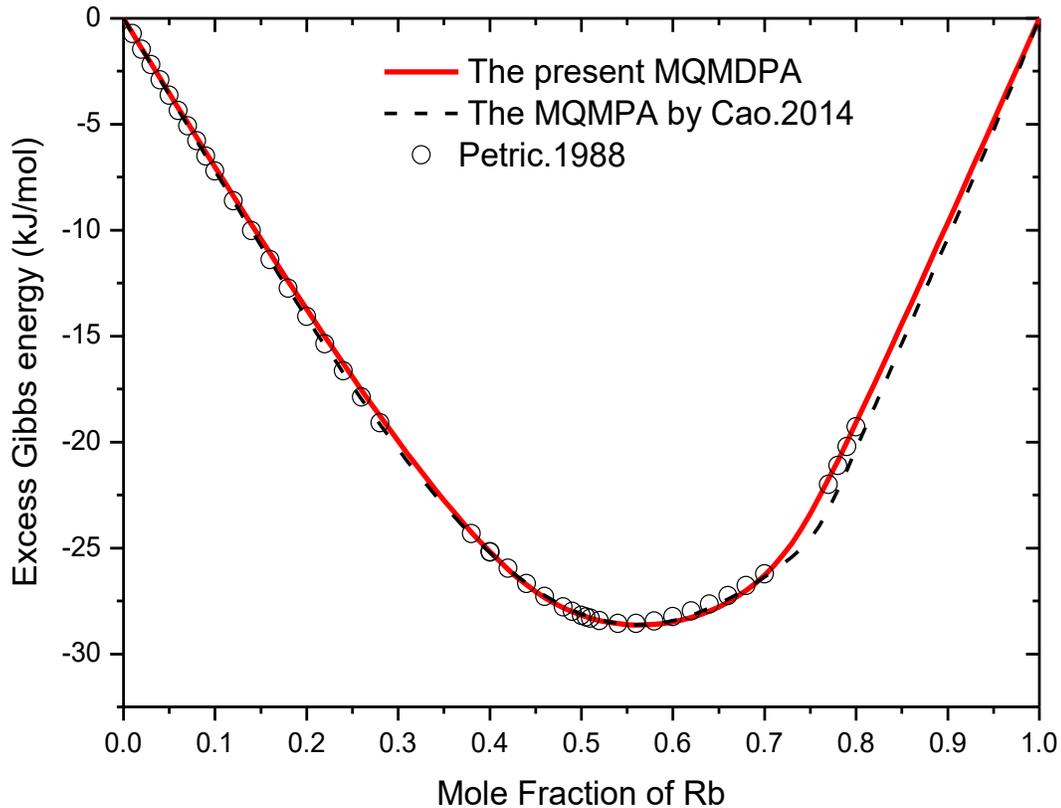

Fig. 9. Calculated Gibbs energy of mixing for the Bi-Rb liquid at 873 K compared with the experimental data [21]

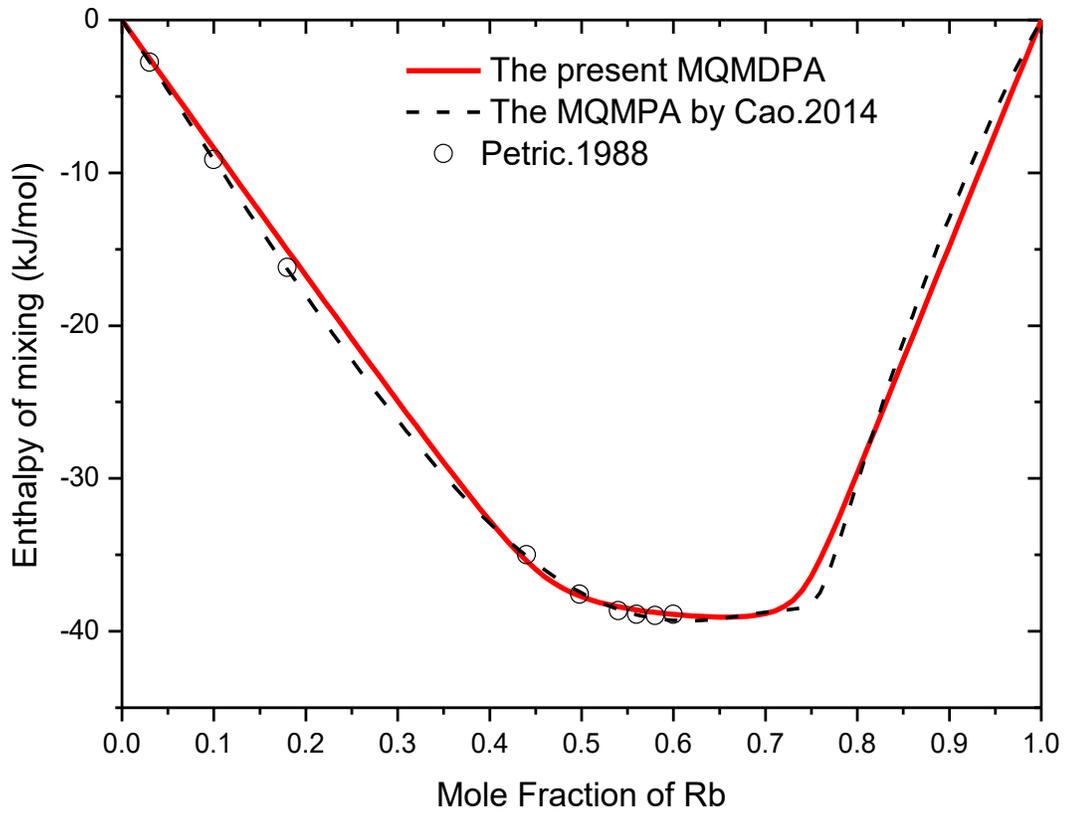

Fig. 10. Calculated enthalpy of mixing for the Bi-Rb liquid at 873 K compared with the experimental data [21]

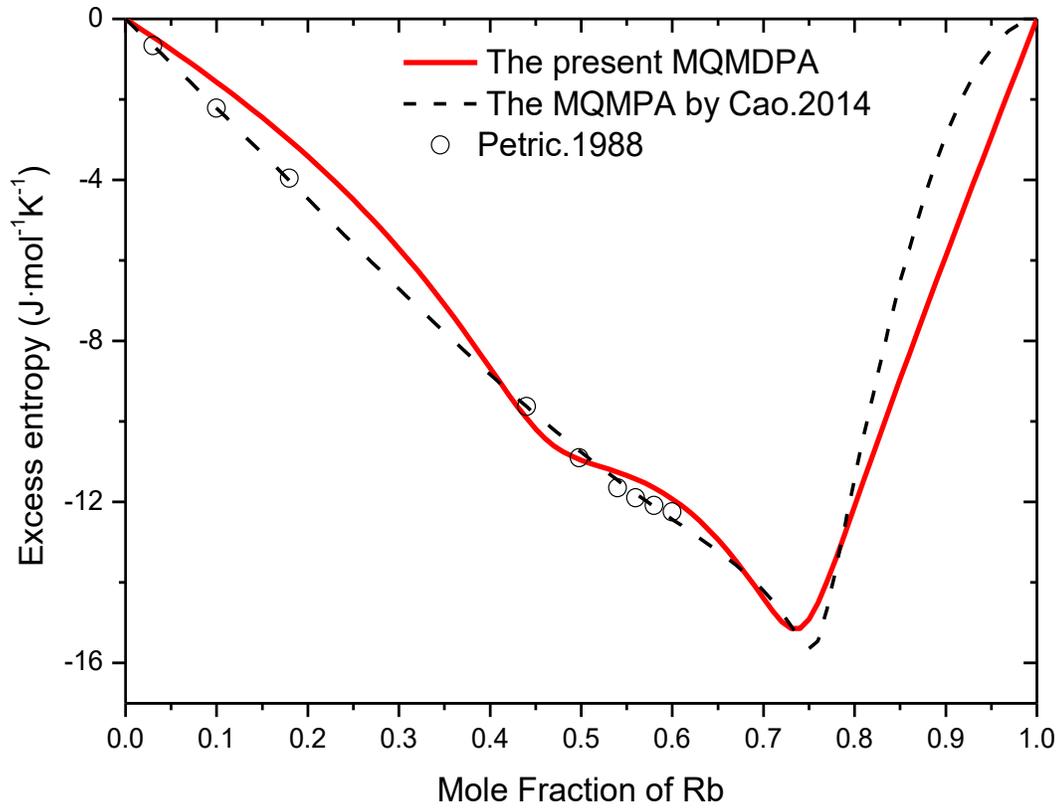

Fig. 11. Calculated excess entropy for the Bi-Rb liquid at 873 K compared with the experimental data [21]

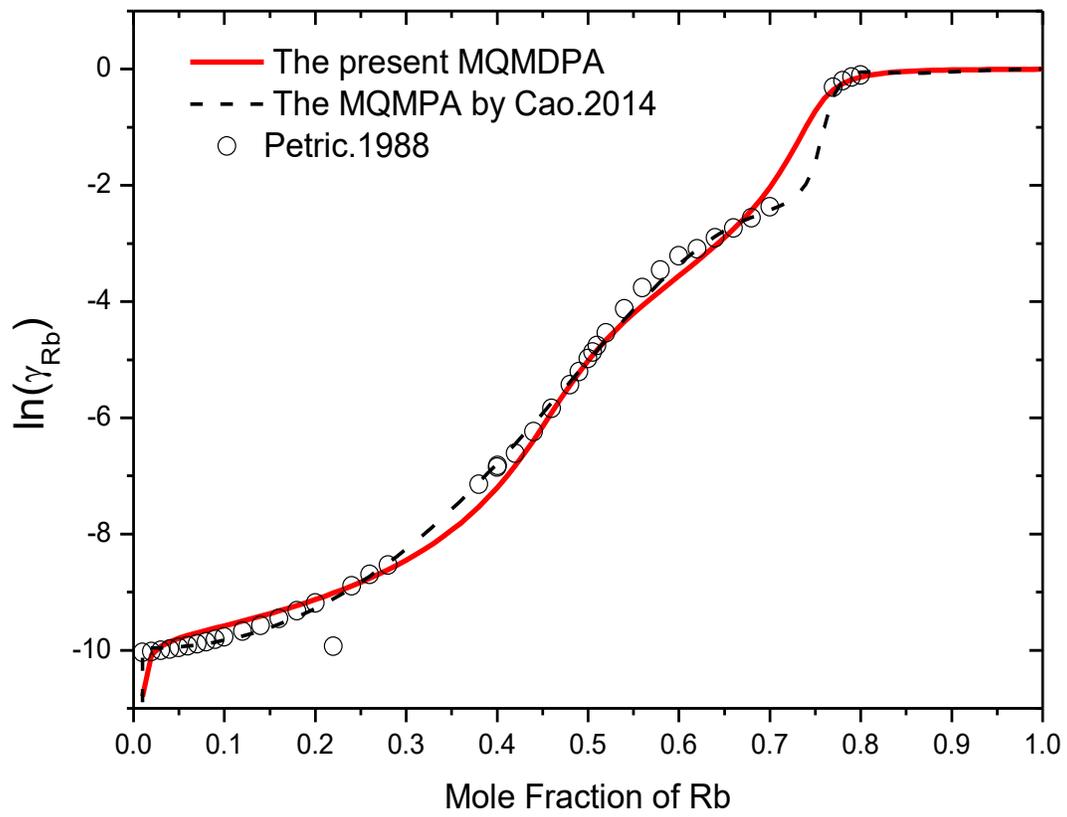

Fig. 12. Calculated activity coefficient of Rb in the Bi-Rb liquid at 873 K compared with the experimental data [21]

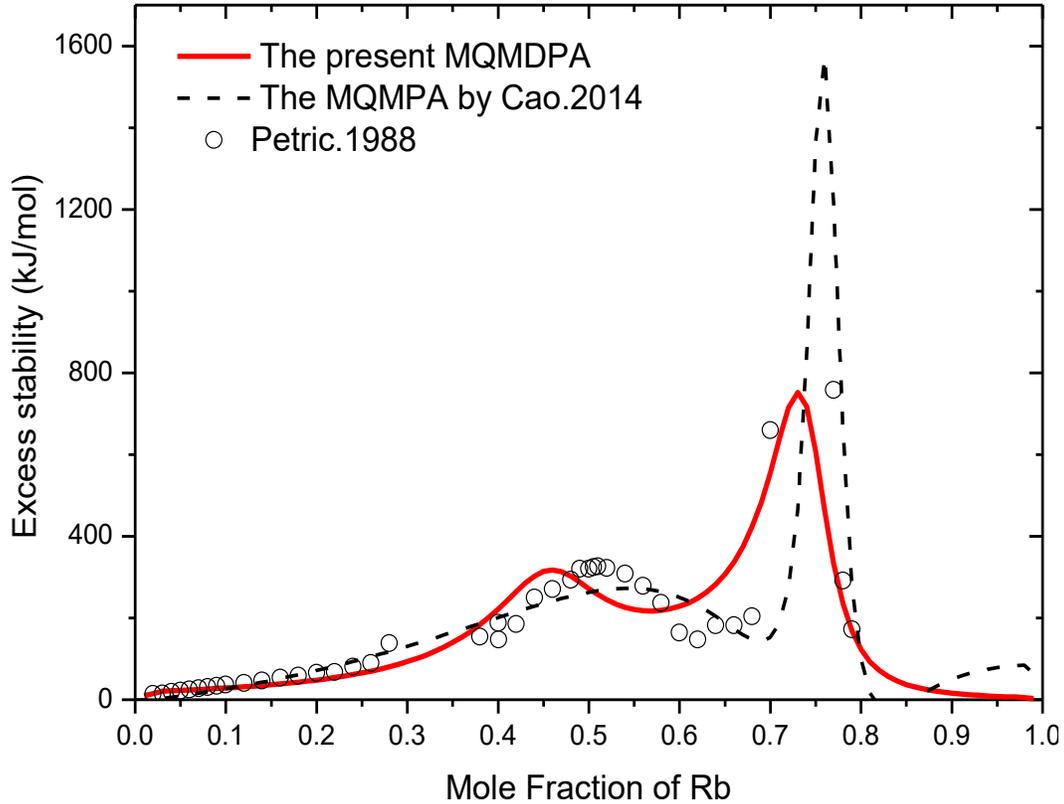

Fig. 13. Calculated excess stability function in the Bi-Rb liquid at 873 K compared with the experimental data [21]

**3.3 The Na-Pb system**

The thermodynamic properties of the Na-Pb liquid were measured from 653 to 753 K and over the entire composition range by EMF measurements [23]. The partial excess Gibbs energy, integral Gibbs energies, enthalpies and entropies of the liquid at 723 K were provided. The calculated ESF curve from the measured scatter data presents three inflections at 50, 66 and 80 at.% Na. The inflections at 50 and 80 at.% Na correspond to maxima, and the inflection at 66 at.% Na corresponds to a minimum, indicating that at least twofold SROs exist. In view of the unusual properties of the Na-Pb liquid, there will most likely still be many model parameters required if the ASM is used with two associates of $Na_4Pb$ and $NaPb$. The inclusion of a large number of model parameters is not conducive to extending them into multicomponent solutions. The present work uses MQMDPA to treat Na-Pb liquid assuming two compositions of maximum SRO with the stoichiometry of $Na_4Pb$ and $NaPb$. In addition to the coordination numbers, four model parameters independent of composition as listed in Table 1, are also used to effectively reproduce all the experimental data.

Figs.14-16 present the calculated Gibbs energy of mixing, enthalpy of mixing, and entropy of mixing for the Na-Pb liquid at 723 K compared with the experimental data [23]. These figures show that the calculated results using the MQMDPA are in good accordance with the experimental data. The calculated activity coefficient of Na versus $X_{Na}$ at 723 K, as shown in Fig. 17, also displays good agreement with the experimental data. The curve for the activity coefficient versus composition exhibits striking and weak steepness at approximately 80 at.% and 50 at.% Na, indicating the formation of strong and weak SROs in the melt, respectively. Twofold SROs can also be clearly observed from the calculated ESF curve, as presented in Fig. 18. Fine tuning of the model parameters to drive the calculated peaks to approach the experimental data is still meaningless owing to the same reasons discussed above.

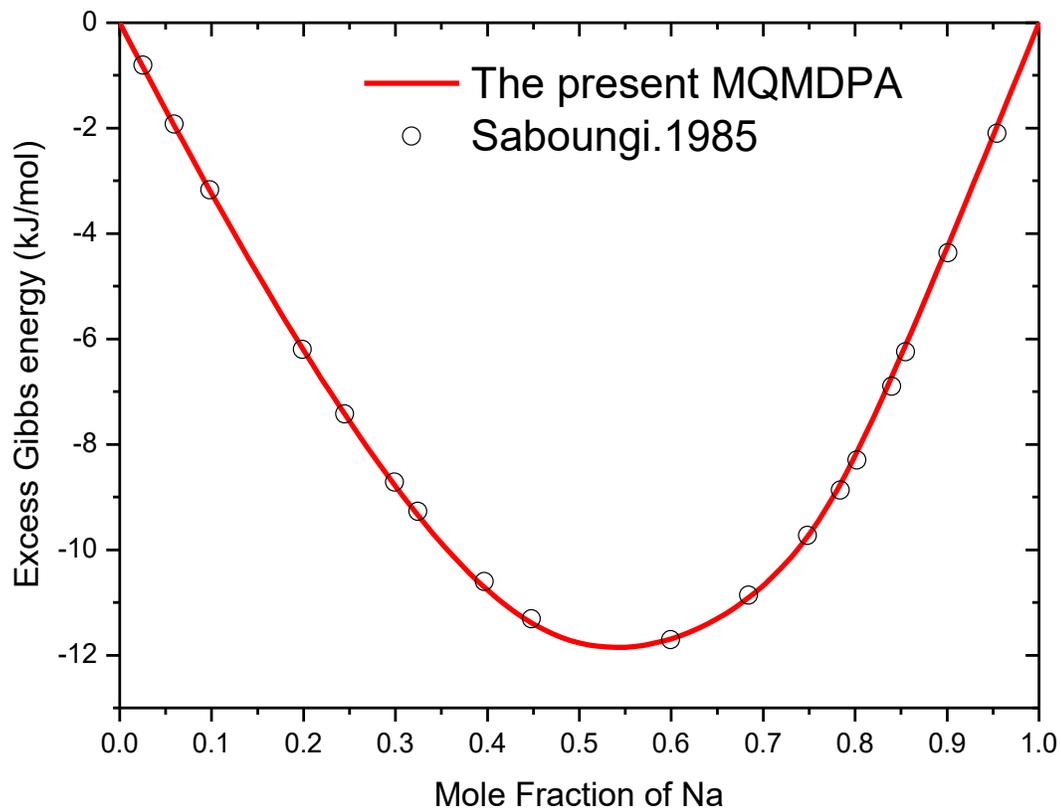

Fig. 14. Calculated Gibbs energy of mixing for the Na-Pb liquid at 723 K compared with the experimental data [23]

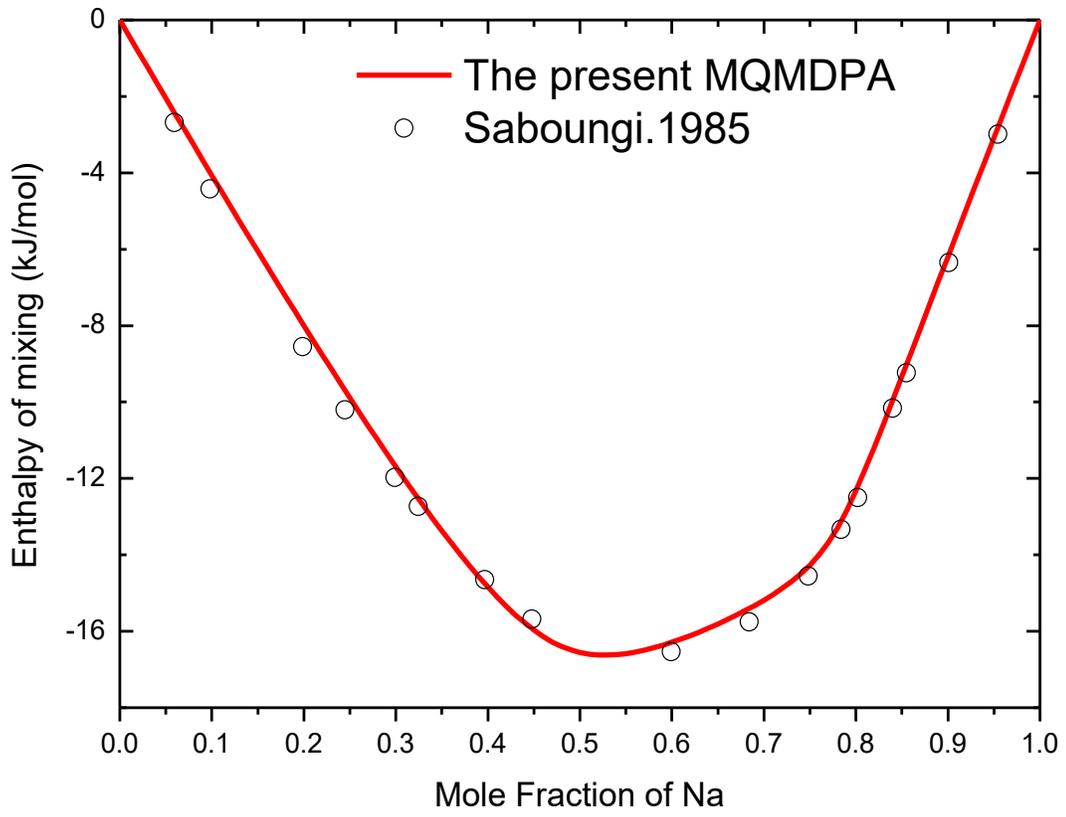

Fig. 15. Calculated enthalpy of mixing for the Na-Pb liquid at 723 K compared with the experimental data [23]

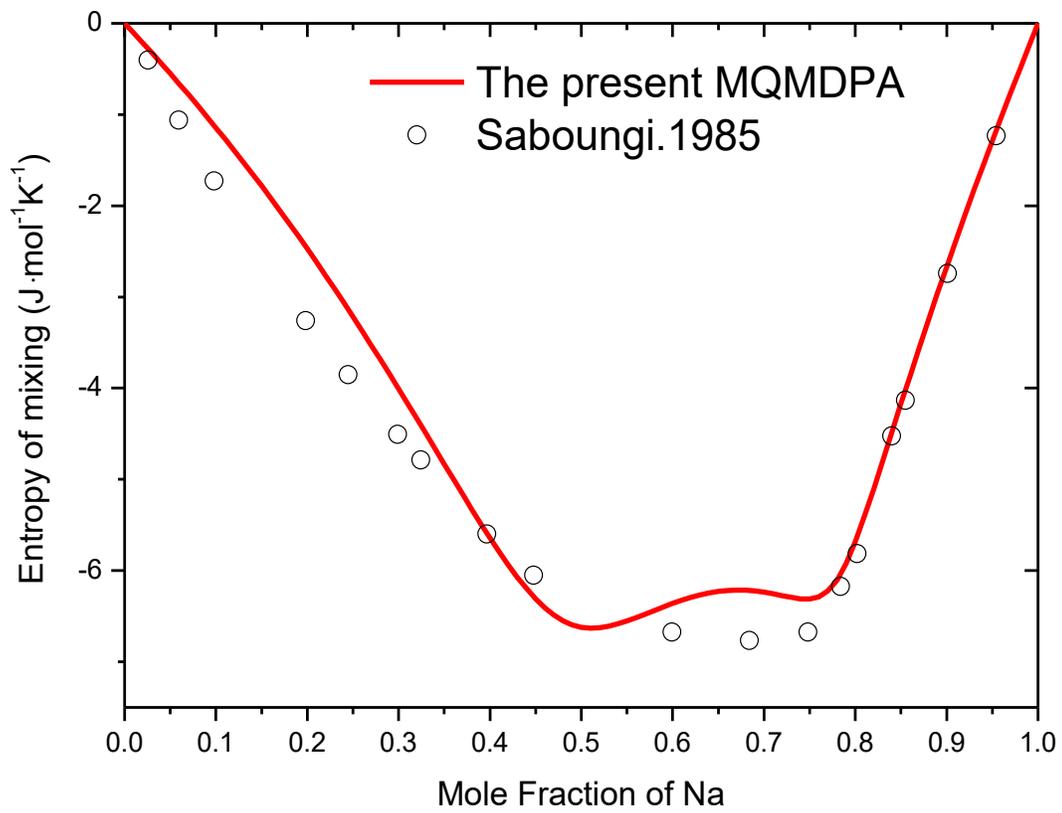

Fig. 16. Calculated entropy of mixing for the Na-Pb liquid at 723 K compared with the experimental data [23]

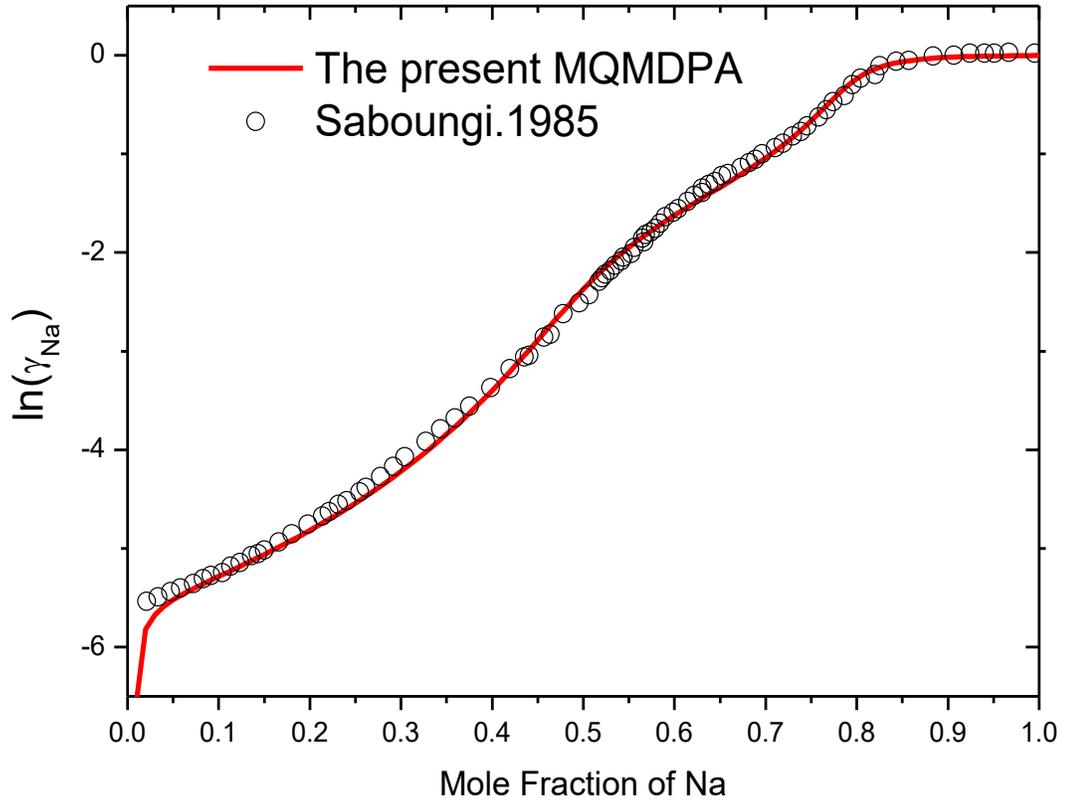

Fig. 17. Calculated activity coefficient of Na in the Na-Pb liquid at 723 K compared with the experimental data [23]

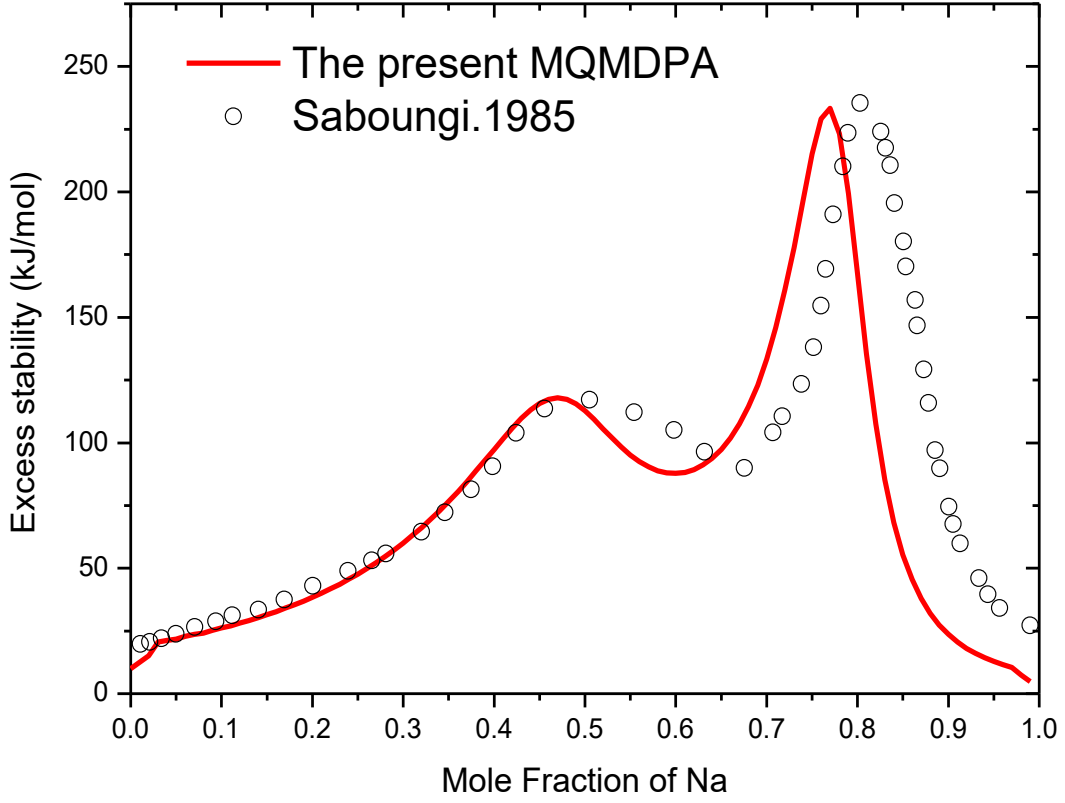

Fig. 18. Calculated excess stability function in the Na-Pb liquid at 723 K compared with the experimental data [23]

## 4. Discussion

In this section, the Modified Quasichemical Model in the Indistinguishable-Pair Approximation (MQMIPA) is proposed, the widely used solution models are reviewed, and their comparisons with the MQMDPA are presented, as well as their pros and cons in the thermodynamic description of solutions.

**4.1 The MQMIPA**

Assuming that there is a trivial difference in distance and coordination among the A-B pairs, the indistinguishable aspect of the A-B pairs is thus emphasized. The A-B pairs are still separated into $Q$ groups owing to the negligible distinctness. In each group, the A-B pairs have completely identical bond energy and coordination numbers. With this assumption, Equation (4) is modified as

$$\Delta S^{pair} = -R\left[n_{AA}\ln\left(\frac{X_{AA}}{Y_A^2}\right) + n_{BB}\ln\left(\frac{X_{BB}}{Y_B^2}\right) + \left(\sum_{k=1}^{Q} n_{AB}^{\langle k\rangle}\right)\ln\left(\frac{\sum_{k=1}^{Q} X_{AB}^{\langle k\rangle}}{2Y_A Y_B}\right)\right] \quad (37)$$

Combining equations (1-3, 37) gives,

$$G = n_A g_A^0 + n_B g_B^0 + RT(n_A \ln X_A + n_B \ln X_B) + RT\left[n_{AA} \ln\left(\frac{X_{AA}}{Y_A^2}\right) + n_{BB} \ln\left(\frac{X_{BB}}{Y_B^2}\right) + \left(\sum_{k=1}^{Q} n_{AB}^{\langle k \rangle}\right) \ln\left(\frac{\sum_{k=1}^{Q} X_{AB}^{\langle k \rangle}}{2Y_A Y_B}\right)\right] + \sum_{k=1}^{Q} n_{AB}^{\langle k \rangle} \left(\frac{\Delta g_{AB}^{\langle k \rangle}}{2}\right) \quad (38)$$

where all the variables and their interrelations are identically defined as equations (6-18). Following the same solution procedure as Equation (19), the equilibrium constant for the quasichemcial reaction is derived as

$$\left(\frac{X_{AA}}{Y_A^2}\right)^{C_{AA}^{\langle k \rangle}} \left(\frac{X_{BB}}{Y_B^2}\right)^{C_{BB}^{\langle k \rangle}} \left(\frac{\sum_{k=1}^{Q} X_{AB}^{\langle k \rangle}}{2Y_A Y_B}\right) = \exp\left(-\frac{\Delta g_{AB}^{\langle k \rangle}}{2RT}\right) \quad (39)$$

where $C_{AA}^{\langle k \rangle}$ and $C_{BB}^{\langle k \rangle}$ have been defined in equations (21-22). Subjected to the mass-balance constraints of equations (17-18), the internal variables $n_{AA}$, $n_{BB}$ and $n_{AB}^{\langle k \rangle}$ are calculated by assigning values to various $\Delta g_{AB}^{\langle k \rangle}$ and then substituting them into Equation (38) to obtain the Gibbs energy of the A-B solution at the constant $n_A$ and $n_B$.

According to Equation (39), the MQMIPA can also reach the three limiting cases. When all the $\Delta g_{AB}^{\langle k \rangle}$ values approach zero, the Bragg-Williams approximation is realized for the first limiting case because $X_{AA} = Y_A^2$, $X_{BB} = Y_B^2$ and $\sum_{k=1}^{Q} X_{AB}^{\langle k \rangle} = 2Y_A Y_B$. When one of the $\Delta g_{AB}^{\langle k \rangle}$ values is much more negative, a onefold SRO is formed for the second limiting case since just one of the $n_{AB}^{\langle k \rangle}$ values is nonzero. When all the A-B pairs have the same bond energy and coordination numbers, the MQMPA is realized for the last limiting case because all the $n_{AB}^{\langle k \rangle}$ values are equal.

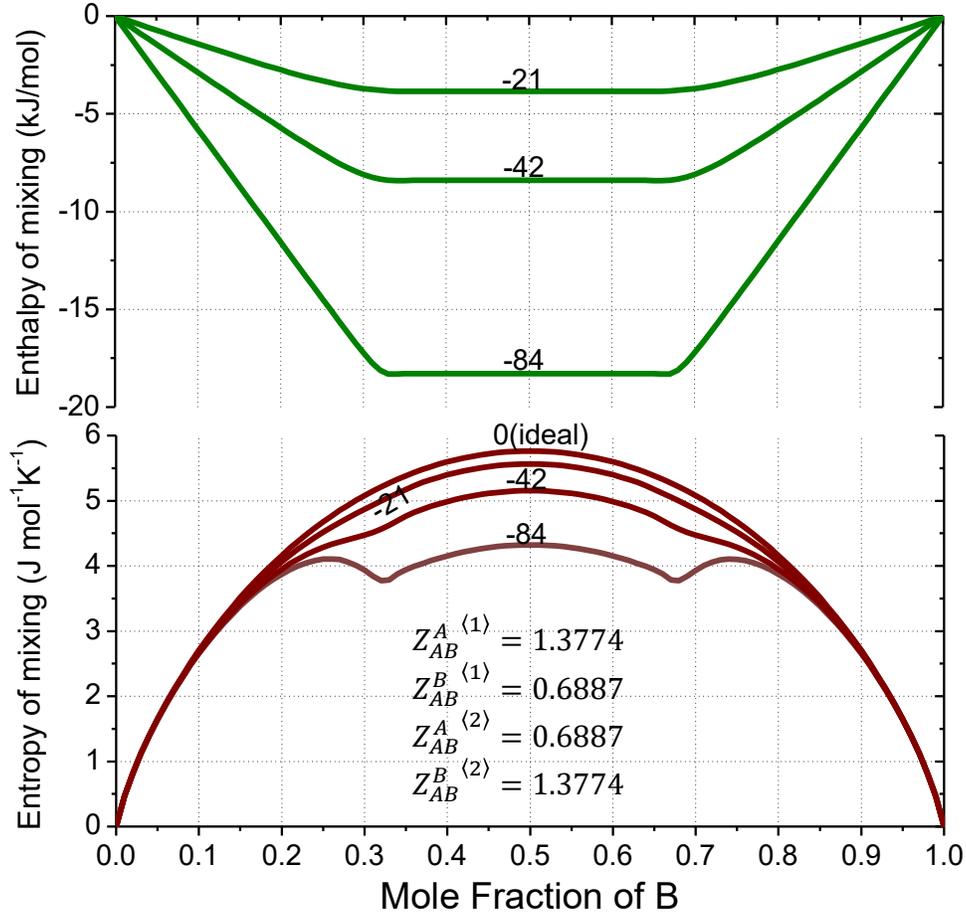

Fig. 19 Molar enthalpy and entropy of mixing for an arbitrary A-B system calculated at 1000 °C from MQMIPA for twofold SROs with $\Delta g_{AB}^{\langle 1 \rangle} = \Delta g_{AB}^{\langle 2 \rangle} = 0, -21, -42$ and -84 kJ/mol

However, the MQMIPA encounters a serious difficulty in the thermodynamic description of solutions when all the $\Delta g_{AB}^{\langle k \rangle}$ values become negative. Taking the two SROs to be captured as examples, the slope of the configurational entropy is discontinuous around the two SRO compositions, and the enthalpy of the mixing curve is too flat, as illustrated in Fig. 19. These problems result from the immiscibility of the two groups of the A-B pairs and cannot be eliminated even when the $\Delta g_{AB}^{\langle k \rangle}$ values are not as negative. The immanent factor is related to the self-contradictory parts used to assemble the MQMIDA. Configurational entropy is proposed based on the concept of identical A-B pairs, which should have unique bond energy and coordination numbers. If either the bond energies or coordination numbers are different for the A-B pairs in different groups, they become distinguishable. As a result, MQMIDA can only be used to describe completely identical pairs (reduced to MQMPA) and cannot be compatible with distinguishable A-B pairs. Therefore, the MQMIDA is not feasible for treating solutions with multiple SRO compositions. This overturns the original conjecture that manifold SROs in binary solutions are related to the identical ordered bonds, which exhibit two minima on the bond-energy curve, and of which energies fluctuate with respect to composition.

### 4.2 The combinatorial model

The MQMPA coupled with the MSM can also provide different types of ordered pairs to treat manifold SROs in binary solutions. For the A-B solution, if pure B is treated by the MSM with two states ($B'$ and $B''$), the combinatorial model should take into account three quasichemical reactions,

$$(A - A) + (B' - B') = 2(A - B') \qquad \Delta g_{AB'} \qquad (40)$$

$$(A - A) + (B'' - B'') = 2(A - B'') \qquad \Delta g_{AB''} \qquad (41)$$

$$(B' - B') + (B'' - B'') = 2(B' - B'') \qquad \Delta g_{B'B''} \qquad (42)$$

where reactions (40-41) are responsible for the two SROs in the solution while reaction (42) occurs in the hypothetical B solution. By randomly distributing all six pairs, the expression of the combinatorial model is then given as

$$G = n_A g_A^o + n_{B'} g_{B'}^o + n_{B''} g_{B''}^o + RT(n_A \ln X_A + n_{B'} \ln X_{B'} + n_{B''} \ln X_{B''}) + RT(n_{AA} \ln \frac{X_{AA}}{Y_A^2}$$
$$+ n_{B'B'} \ln \frac{X_{B'B'}}{Y_{B'}^2} + n_{B''B''} \ln \frac{X_{B''B''}}{Y_{B''}^2} + n_{AB'} \ln \frac{X_{AB'}}{2Y_A Y_{B'}} + n_{AB''} \ln \frac{n_{AB''}}{2Y_A Y_{B''}} + n_{B'B''} \ln \frac{n_{B'B''}}{2Y_{B'} Y_{B''}})$$
$$+ (n_{AB'} \frac{\Delta g_{AB'}}{2} + n_{AB''} \frac{\Delta g_{AB''}}{2} + n_{B'B''} \frac{\Delta g_{B'B''}}{2}) \qquad (43)$$

where all the variables and their interrelations have definitions similar to those in equations (7-18). There exists an additional constraint of $n_B = n_{B'} + n_{B''}$ to calculate the equilibrium distribution of pairs and the Gibbs energy of the A-B solution. $\Delta g_{AB'}$ and $\Delta g_{AB''}$ are the major factors controlling the existence of the two SROs in the A-B solution. $\Delta g_{B'B''}$ seems to be irrespective of the twofold SROs, whereas it also influences their existence. Another influencing factor is $dg = g_{B''}^o - g_{B'}^o$, which characterizes the stability of the hypothetical $B''$ state in the B solution. By leveraging $\Delta g_{AB'}$, $\Delta g_{AB''}$, $\Delta g_{B'B''}$ and $dg$, the combinatorial model can capture twofold SROs in the A-B solution. By setting $\Delta g_{AB'} = \Delta g_{AB''} = \Delta g_{B'B''} = 0$ and $dg = \infty$, the combinatorial model can be reduced to the BWM. By assigning $\Delta g_{AB'}$ a more negative value and setting $\Delta g_{AB''} = \Delta g_{B'B''} = 0$ and $dg = \infty$, the combinatorial model can reach the limit where onefold SRO appears in the calculated results. However, in any case, the combinatorial model cannot be completely transformed to the MQMPA.

In practical applications, assignments of the $\Delta g_{B'B''}$ and $dg$ values are actually very arbitrary and lack physical foundations. One requirement for defining $\Delta g_{B'B''}$ and dg is inability to change the melting point of pure B. Normally, $\Delta g_{B'B''}$ is set to zero for simplicity. For the hypothetical B solution, Equation (43) reduces to the MSM expression,

$$G_B = (1 - n_{B''})g^o_{B'} + n_{B''}g^o_{B''} + RT(n_{B''} \ln X_{B''} + (1 - n_{B''}) \ln(1 - X_{B''})) \quad (44)$$

where $n_{B'} + n_{B''}$ is assumed to be one mole. By setting $\partial G/\partial n_{B''} = 0$, the equilibrium $n_{B''}$ can be calculated as,

$$n_{B''} = \frac{1}{1+\exp\left(\frac{dg}{RT}\right)} \quad (45)$$

Substituting Equation (45) into Equation (44) gives

$$G_B = g^o_{B'} - RT\ln\left(1 + \exp\left(-\frac{dg}{RT}\right)\right) \quad (46)$$

From Equation (46), the Gibbs energy of the hypothetical B solution is equivalent to that of the pure B component when $dg$ approaches positive infinity. If $dg$ is arbitrarily assigned as 100 kJ/mol, $n_{B''}$ and $G_B - g^o_{B'}$ can be calculated to be $7.8849 \times 10^{-5}$ mole and -0.8351 J/mol at 1000 °C via Equations (45-46), respectively. This hardly changes the melting point of the B component.

The combinatorial model actually employs the energy formalism for the pseudoternary solution to mimic the thermodynamic behavior of the binary solution with two SROs. If $\Delta g_{AB'}$ and $\Delta g_{AB''}$ are composition dependent, a geometrical solution model is required to interpolate the pseudoternary Gibbs energy from the sub-binaries. As shown in Fig.20, there are two types of interpolation models. If binary $\Delta g_{AB'}$ is expressed as Equation (6), the symmetrical interpolation model gives

$$\Delta g_{AB'} = \Delta g^o_{AB'} + \sum_{i+j\geq 1} g^{ij}_{AB'} \left(\frac{X_{AA}}{X_{AA}+X_{AB'}+X_{B'B'}}\right)^i \left(\frac{X_{B'B'}}{X_{AA}+X_{AB'}+X_{B'B'}}\right)^j \quad (47)$$

while the asymmetrical interpolation model generates

$$\Delta g_{AB'} = \Delta g^o_{AB'} + \sum_{i+j\geq 1} g^{ij}_{AB'} X_{AA}{}^i (X_{B'B'} + X_{B'B''} + X_{B''B''})^j \quad (48)$$

for use in the ternary solution [2]. Similar methods can be employed to interpolate binary $\Delta g_{AB''}$ for use in the ternary solution. Hence, there are four completely different ways to interpolate the Gibbs energy of the pseudoternary solution. It is thus uncertain which one is the most accurate way to predict the thermodynamic properties of the binary A-B solution.

The combinatorial model hardly avoids the aforementioned uncertainties of how to physically select the best geometric solution model and how to physically provide clear definitions of the model parameters. More uncertainties may be encountered when the combinatorial model is used to treat manifold SROs in binary solutions. These uncertainties have adversely impacted the predictive power of the combinatorial model toward those unknown regions in multicomponent systems. However, none of these uncertainties appears in the MQMDPA, since it is unnecessary to couple the MSM and can directly capture manifold SROs in binary solutions. The MQMDPA can thus be safely extended for reliable use in multicomponent solutions. In addition, the calculation efficiency is lower in the combinatorial model than in the MQMDPA, since the former must solve more internal variables. The situation becomes even worse when the combinatorial model is used to treat multicomponent solutions in which each sub-binary contains manifold SROs.

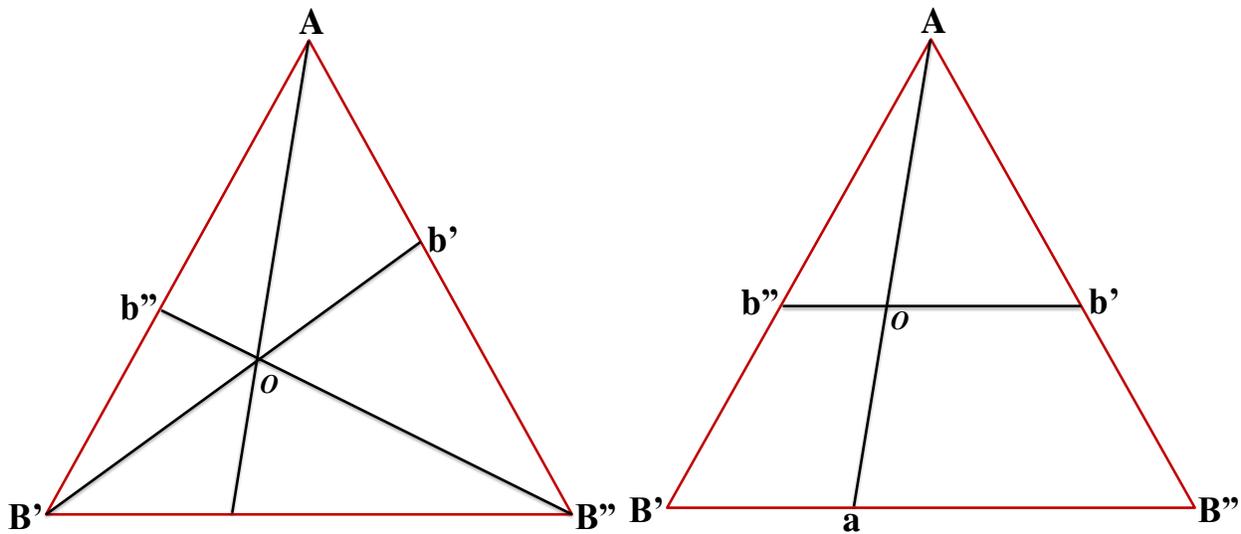

Fig. 20. Geometric interpolation models used for the pseudoternary solution: (a) the symmetrical model of the Kohler type; (b) the asymmetrical model of the Kohler-Toop type

### 4.3 The ASM and ILM

The ASM and ILM are also able to capture manifold SROs in binary solutions. The two models are actually identical and can be transformed between each other if there are no reciprocal terms. Taking the CaO-SiO$_2$ melt as an example, if the melt has two SROs nearly occurring at Ca$_2$SiO$_4$ and Ca$_3$Si$_2$O$_7$, the ILM with $(Ca^{+2})_P(O^{-2}, SiO_4^{-4}, Si_2O_7^{-6}, SiO_2^0)_Q$ can be used to describe its thermodynamic properties. The Gibbs energy of the melt is then given,

$$G = y_{O^{-2}}g^0_{Ca^{+2}:O^{-2}} + y_{SiO_4^{-4}}g^0_{Ca^{+2}:SiO_4^{-4}} + y_{Si_2O_7^{-6}}g^0_{Ca^{+2}:Si_2O_7^{-6}} + Qy_{SiO_2^0}g^0_{SiO_2^0} +$$
$$QRT(y_{O^{-2}}\ln y_{O^{-2}} + y_{SiO_4^{-4}}\ln y_{SiO_4^{-4}} + y_{Si_2O_7^{-6}}\ln y_{Si_2O_7^{-6}} + y_{SiO_2^0}\ln y_{SiO_2^0}) +$$
$$g^E(y_{O^{-2}}, y_{SiO_4^{-4}}, y_{Si_2O_7^{-6}}, y_{SiO_2^0}) \tag{49}$$

where $Q = 1$ and $P = y_{O^{-2}} + 2y_{SiO_4^{-4}} + 3y_{Si_2O_7^{-6}}$ should be maintained for the electroneutral condition, and $g^E$ represents the excess Gibbs energy and is usually written as a polynomial function of the ion fractions. Owing to $Ca^{+2}$ fully occupying the cationic sublattice, the anion fractions $y_{O^{-2}}, y_{SiO_4^{-4}}, y_{Si_2O_7^{-6}}$ and $y_{SiO_2^0}$ in the anionic sublattice can be replaced by the species fractions $y_{CaO}, y_{Ca_2SiO_4}, y_{Ca_3Si_2O_7}$ and $y_{SiO_2^0}$ in the melt, respectively. $g^0_{Ca^{+2}:O^{-2}}$ stands for the standard molar Gibbs energy of CaO, $g^0_{Ca^{+2}:SiO_4^{-4}}$ for $Ca_2SiO_4$, $g^0_{Ca^{+2}:Si_2O_7^{-6}}$ for $Ca_3Si_2O_7$ and $g^0_{SiO_2^0}$ for $SiO_2$. Equation (49) can then be transformed to the ASM expression,

$$G = y_{CaO}g^0_{CaO} + y_{Ca_2SiO_4}g^0_{Ca_2SiO_4} + y_{Ca_3Si_2O_7}g^0_{Ca_3Si_2O_7} + y_{SiO_2^0}g^0_{SiO_2^0} + RT(y_{CaO}\ln y_{CaO}$$
$$+ y_{Ca_2SiO_4}\ln y_{Ca_2SiO_4} + y_{Ca_3Si_2O_7}\ln y_{Ca_3Si_2O_7} + y_{SiO_2^0}\ln y_{SiO_2^0})$$
$$+ g^E(y_{O^{-2}}, y_{SiO_4^{-4}}, y_{Si_2O_7^{-6}}, y_{SiO_2^0}) \tag{50}$$

where the species fractions and the melt compositions have the following mass-balance relations:

$$n_{CaO} = y_{CaO} + 2y_{Ca_2SiO_4} + 3y_{Ca_3Si_2O_7} \tag{51}$$
$$n_{SiO_2^0} = y_{Ca_2SiO_4} + 2y_{Ca_3Si_2O_7} + y_{SiO_2^0} \tag{52}$$

$g^0_{Ca_2SiO_4}$ and $g^0_{Ca_3Si_2O_7}$ are usually defined through the association reactions,

$$2CaO + SiO_2 = Ca_2SiO_4 \quad dg^0_{Ca_2SiO_4} \tag{53}$$
$$2CaO + 3SiO_2 = Ca_3Si_2O_7 \quad dg^0_{Ca_3Si_2O_7} \tag{54}$$

where

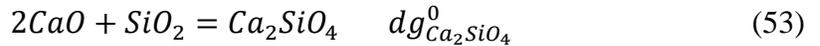
$$dg^0_{Ca_2SiO_4} = g^0_{Ca_2SiO_4} - 2g^0_{CaO} - g^0_{SiO_2^0} = -RT\ln K_{Ca_2SiO_4} \tag{55}$$
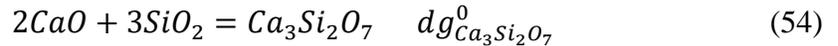
$$dg^0_{Ca_3Si_2O_7} = g^0_{Ca_3Si_2O_7} - 2g^0_{CaO} - 3g^0_{SiO_2^0} = -RT\ln K_{Ca_3Si_2O_7} \tag{56}$$

If $g^E = 0$, then Equation (50) is identical to that of a gas. Otherwise, Equation (50) is identical to the expression commonly used for the Gibbs energy of a quaternary $CaO$-$SiO_2$-$Ca_2SiO_4$-$Ca_3Si_2O_7$ system with an explicit polynomial expression for $g^E$.

The ASM and ILM are unable to reach the first limiting case of reducing to the BWM with the ideal entropy of mixing. When $dg^0_{Ca_2SiO_4}$ and $dg^0_{Ca_3Si_2O_7}$ of reactions (53-54) are equal to zero, then $K_{Ca_2SiO_4} = K_{Ca_3Si_2O_7} = 1$, and the equilibrium numbers of the associates $Ca_2SiO_4$ and Ca$_3$Si$_2$O$_7$ are not zero. The numbers of associates are only zero when $dg^0_{Ca_2SiO_4}$ and $dg^0_{Ca_3Si_2O_7}$ approach positive infinity. This is the so-called entropy paradox pointed out by Lück et al. [24]. The ASM and ILM can reach the second limiting case where only one of the manifold SROs exists, and the others almost disappear when the corresponding association is assigned a more negative formation energy. The ASM and ILM can reach the third limiting case where manifold SROs can merge to form onefold SROs in binary solutions if all the associates possess the same stoichiometry and Gibbs energy. The primary difference between the MQMPDA and the ASM and ILM is that the former model can reach the first limiting case while the latter two fail in this respect. The reason will be given by the following comparison between the MQMDPA and the ASM expressions.

The MQMDPA can be expressed in a form very similar to the ASM equation. Considering that there are $Q$ compositions of maximum SRO in the A-B solution, the $Q+2$ associates, $A_{\frac{1}{Z^A_{AA}}}A_{\frac{1}{Z^A_{AA}}}$, $B_{\frac{1}{Z^B_{BB}}}B_{\frac{1}{Z^B_{BB}}}$ and $A_{\frac{1}{Z^{A\langle k\rangle}_{AB}}}B_{\frac{1}{Z^{B\langle k\rangle}_{AB}}}$ (integer $k$ from 1 to $Q$) are thus contained. Let the numbers of moles of the $Q+2$ species be equal to the numbers of moles of the pairs. This gives

$$n_{A_{\frac{1}{Z^A_{AA}}}A_{\frac{1}{Z^A_{AA}}}} = n_{AA}; \quad n_{B_{\frac{1}{Z^B_{BB}}}B_{\frac{1}{Z^B_{BB}}}} = n_{BB}; \quad n^{\langle k\rangle}_{AB} = n_{A_{\frac{1}{Z^{A\langle k\rangle}_{AB}}}B_{\frac{1}{Z^{B\langle k\rangle}_{AB}}}} \tag{57}$$

The mass balance equations (17-18) now become

$$n_A = \frac{2n_{A_{\frac{1}{Z^A_{AA}}}A_{\frac{1}{Z^A_{AA}}}}}{Z^A_{AA}} + \sum_{k=1}^{Q} \frac{n_{A_{\frac{1}{Z^{A\langle k\rangle}_{AB}}}B_{\frac{1}{Z^{B\langle k\rangle}_{AB}}}}}{Z^{A\langle k\rangle}_{AB}} \tag{58}$$

$$n_B = \frac{2n_{B_{\frac{1}{Z^B_{BB}}}B_{\frac{1}{Z^B_{BB}}}}}{Z^B_{BB}} + \sum_{k=1}^{Q} \frac{n_{A_{\frac{1}{Z^{A\langle k\rangle}_{AB}}}B_{\frac{1}{Z^{B\langle k\rangle}_{AB}}}}}{Z^{B\langle k\rangle}_{AB}} \tag{59}$$

Equations (58-59) are "true" chemical mass balances in that the numbers of moles of A and B are equal on both sides of the equations as in equations (51-52) for the ASM. Likewise, the standard molar Gibbs energy of the pairs can be defined,

$$g^o_{AA} = \frac{2}{Z^A_{AA}} g^o_A \tag{60}$$

$$g^o_{BB} = \frac{2}{Z^B_{BB}} g^o_B \tag{61}$$

$$g^{\langle k,o\rangle}_{AB} = \frac{1}{Z^{A\langle k\rangle}_{AB}} g^o_A + \frac{1}{Z^{B\langle k\rangle}_{AB}} g^o_B + \frac{\Delta g^{\langle k,o\rangle}_{AB}}{2} \tag{62}$$

where $\Delta g_{AB}^{\langle k,o \rangle}$ from Equation (6) is the constant term in the expansion of $\Delta g_{AB}^{\langle k \rangle}$ and the Gibbs energy change of the quasichemical reaction (5). Substitution of equations (17-18, 60-62) into Equation (1) and rearrangement of terms then gives

$$G = n_{AA} g_{AA}^o + n_{BB} g_{BB}^o + \sum_{k=1}^{Q} n_{AB}^{\langle k \rangle} g_{AB}^{\langle k,o \rangle} + RT(n_{AA} \ln X_{AA} + n_{BB} \ln X_{BB} + \sum_{k=1}^{Q} n_{AB}^{\langle k \rangle} \ln X_{AB}^{\langle k \rangle})$$

$$+ RT(n_A \ln X_A + n_B \ln X_B - n_{AA} \ln Y_A^2 - n_{BB} \ln Y_B^2 + \sum_{k=1}^{Q} n_{AB}^{\langle k \rangle} \ln(Q/2Y_A Y_B)) + G^E \quad (63)$$

where

$$G^E = \frac{(n_{AA} + n_{BB} + \sum_{k=1}^{Q} n_{AB}^{\langle k \rangle})}{2} \sum_{k=1}^{Q} [X_{AA} X_{AB}^{\langle k \rangle} \sum_{i \geq 1} g_{AB}^{\langle i0 \rangle} X_{AA}^{i-1} + X_{BB} X_{AB}^{\langle k \rangle} \sum_{j \geq 1} g_{AB}^{\langle 0j \rangle} X_{BB}^{j-1}] \quad (64)$$

Apart from the second configurational entropy term, Equation (63) is of the same form as an ASM equation. It is the second entropy term that resolves the "entropy paradox" of the ASM and allows the MQMDPA to reduce to an ideal solution model in the limit.

The comparison between the MQMDPA and the ASM is also beneficial to help implement the MQMDPA into software. The algorithms and subroutines commonly used for the ASM with polynomial expansions for $G^E$ can thus be used directly for the MQMDPA with the simple addition of the extra entropy term. The extension of the MQMDPA to multicomponent solutions is greatly simplified by this formalism as will be discussed in an accompanying paper [25].

## 5. Concluding remarks

The present paper has developed the MQMDPA to directly capture manifold SROs and their thermodynamic behaviors in binary solutions. The development was conducted by grouping the ordered pairs within the MQMPA into several distinguishable types. Each type of ordered pair possesses unique coordination numbers and bond energy, which are responsible for capturing one of the manifold SROs with the desired composition and strength. Once all the distinguishable ordered pairs are assigned the same coordination numbers and bond energy, they become indistinguishable, and MQMDPA is reduced to MQMPA. This is one of the limiting cases that the MQMDPA can reach. The MQMDPA can reach another limiting case of reducing to the Bragg-Williams Model once the Gibbs energy changes of all pair-exchange reactions are zero. The MQMDPA can reach the last limiting case where only one SRO exists and all the others almost disappear if the corresponding ordered pairs are assigned more negative energy. Reaching these limiting cases enables the MQMDPA to be superior to other solution models.

This was demonstrated by the detailed comparisons among the MQMDPA, MQMIPA, ASM, ILM and the combinatorial model. Eventually, three real liquids (Bi-K, Bi-Rb and Na-Pb) with at least two observed SROs were selected to examine the effectiveness and reliability of the MQMDPA to capture their thermodynamic behaviors. It is manifested that much fewer tunable model parameters were required in the MQMDPA than in other solution models, which allows safer extensions of the parameters to ternary and larger multicomponent systems.

## Acknowledgments

The author Kun Wang would like to thank his wife and daughter for all their great support and is also very grateful for their understanding and consideration of the lack of time spent together with them. Emeritus Professor Arthur D. Pelton is also acknowledged for many enlightening discussions on the distinguishability and indistinguishability among the ordered pairs.